\documentclass[aps,prd,twocolumn,a4paper,superscriptaddress,floatfix]{revtex4-1}
\usepackage{graphicx}
\usepackage{color}
\begin{document}

\newcommand{\be}{\begin{equation}}
\newcommand{\ee}{\end{equation}}
\newcommand{\bq}{\begin{eqnarray}}
\newcommand{\eq}{\end{eqnarray}}
\newcommand{\bsq}{\begin{subequations}}
\newcommand{\esq}{\end{subequations}}
\newcommand{\bc}{\begin{center}}
\newcommand{\ec}{\end{center}}
\newcommand{\JV}[1]{{\color{blue} JV: #1}}

\title{Models for Small-Scale Structure on Cosmic Strings: II. Scaling and its stability}
\author{J. P. P. Vieira}
\email[Electronic address: ]{J.Pinto-Vieira@sussex.ac.uk}
\affiliation{Centro de Astrof\'{\i}sica, Universidade do Porto, Rua das Estrelas, 4150-762 Porto, Portugal}
\affiliation{Department of Physics and Astronomy, University of Sussex, Brighton, BN1 9RH, United Kingdom}
\author{C.J.A.P. Martins}
\email[Electronic address: ]{Carlos.Martins@astro.up.pt}
\affiliation{Centro de Astrof\'{\i}sica, Universidade do Porto, Rua das Estrelas, 4150-762 Porto, Portugal}
\affiliation{Instituto de Astrof\'{\i}sica e Ci\^encias do Espa\c co, CAUP, Rua das Estrelas, 4150-762 Porto, Portugal}
\author{E.P.S. Shellard}
\email[Electronic address: ]{E.P.S.Shellard@damtp.cam.ac.uk}
\affiliation{Department of Applied Mathematics and Theoretical Physics, Centre for Mathematical Sciences,\\ University of Cambridge, Wilberforce Road, Cambridge CB3 0WA, United Kingdom}

\date{6 August 2016}
\begin{abstract}
We make use of the formalism described in a previous paper [Martins {\it et al.} Phys. Rev. D90 (2014) 043518] to address general features of wiggly cosmic string evolution. In particular, we highlight the important role played by poorly understood energy loss mechanisms and propose a simple ansatz which tackles this problem in the context of an extended velocity-dependent one-scale model. We find a general procedure to determine all the scaling solutions admitted by a specific string model and study their stability, enabling a detailed comparison with future numerical simulations. A simpler comparison with previous Goto-Nambu simulations supports earlier evidence that scaling is easier to achieve in the matter era than in the radiation era. In addition, we also find that the requirement that a scaling regime be stable seems to notably constrain the allowed range of energy loss parameters.
\end{abstract}
\maketitle

\section{\label{sint}Introduction}

Vortex-lines or topological strings are ubiquitous in physical contexts, with perhaps the most interesting and well-studied examples being cosmic strings in the early universe and vortex-lines in superfluid helium. (For extensive reviews on the subject see \cite{VSH,COND1,COND2,GEYER}.) Their nonlinear nature and interactions imply that the detailed quantitative understanding of their properties and experimental or observational consequences is a significant challenge, which is compounded by the the complexity of evolving a full network. This is particularly topical given the recent availability of high-quality data which one may use to constrain these models, such as that of the Planck satellite \cite{Planck}. In the future, gravitational waves should become an additional observational window \cite{DamourVilenkin}.

A significant part of this effort must therefore be based on numerical simulations, but these are both technically difficult and very computationally costly \cite{BB,AS,FRAC,RSB,VVO,Stuckey,Blanco,Hiramatsu}. This is among the motivations for developing complementary analytic approaches, essentially abandoning the detailed \textit{statistical physics} of the string network to concentrate on its \textit{thermodynamics}. For the simplest Goto-Nambu string networks, which have been the subject of most studies so far, the velocity-dependent one-scale (VOS) model \cite{MS1,MS2,MS3,MS4} has been exhaustively studied, and its quantitative success has been extensively demonstrated by direct comparison with both field theory and Goto-Nambu numerical simulations \cite{FRAC,ABELIAN}. The model allows one to describe the scaling laws and large-scale properties of string networks in both cosmological and condensed matter settings with a minimal number of free parameters. More elaborate approaches have also been introduced with the goal of explicitly describing the behavior of small-scale structures on the strings \cite{ACK,POLR}. Here we study a similar extension for the VOS model.

Indeed, cosmologically realistic string networks are not expected to be of Goto-Nambu type. In particular, the previously mentioned simulations of cosmic strings in expanding universes have established beyond doubt the existence of a significant amount of short-wavelength propagation modes (commonly called \textit{wiggles}) on the strings, on scales that can be several orders of magnitude smaller than the correlation length. In a previous paper \cite{PAP1} we introduced a mathematical formalism suitable for the description of the evolution of both large-scale and small-scale properties of a cosmic string network in expanding space. In particular, we arrived at a complete set of equations which allows us to model the evolution of such important quantities as the characteristic length of the network, a characteristic velocity, and both the multifractal dimension and the effective energy per unit length of the strings. There the focus of the applications was on two simplified limits of physical relevance: the tensionless and the linear limit (the latter being especially appropriate for comparison with Abelian-Higgs network simulations). 

This paper continues the exploration of this formalism. After a brief overview of the main results of the first paper, we focus our attention on a general study of the scaling regimes allowed by this model, including their attractor behavior. These results will be illustrated for the case of a simple ansatz which naturally generalizes the energy loss mechanisms considered in the simpler one-scale-type models. Finally, we use our results to make a first comparison with previously existing numerical simulations. A more detailed comparison will require new simulations (both because additional diagnostics should be output and because a higher resolution would be desirable) and is left for subsequent work.

\section{\label{elast}Elastic String Evolution}

The VOS model \cite{PHD,MS2,MS1} is the simplest and most reliable method for calculating the evolution of the large-scale properties of a network of Goto-Nambu cosmic strings obeying the action
\begin{equation}
S=-\mu_{0}\int\sqrt{-\gamma}d^{2}\sigma\label{eq:GN_action}
\end{equation}
where $\sigma^{a}$ are the string worldsheet coordinates, $\gamma$ is the determinant of $\gamma^{ab}$, the pullback metric on the worldsheet, and $\mu_{0}$ is the string mass per unit length (equal to the local string tension) which is generally expected to be of the order of the square of the symmetry breaking scale associated with the formation of the strings. At the expense of assuming there is only one relevant length scale $L$ in the network (as in Kibble's one-scale model \cite{KIB}), this model allows us to make quantitative predictions about the evolution of the energy in the network $E$ as well as a RMS velocity $v$ defined by
\begin{equation}
E=\mu_{0}a\int\epsilon d\sigma\propto\frac{\mu_{0}a^{3}}{L^{2}},\,\quad v^{2}=\frac{\int\mathbf{\dot{x}}^{2}\epsilon d\sigma}{\int\epsilon d\sigma}\label{eq:L_v}
\end{equation}
where $a$ is the scale factor of an FLRW metric
\begin{equation}
ds^{2}=a^{2}\left(d\tau^{2}-d\mathbf{x}^{2}\right)\label{FRW}
\end{equation}
In particular, it is found that if the scale factor behaves as a power law of the form
\begin{equation}
a\propto t^{\lambda}\label{a_power}
\end{equation}
where $\lambda$ is a constant between $0$ and $1$, then there is an attractor \textit{scaling} regime defined by $L/t=const.$ and $v=const$.

Throughout this discussion, our aim is to emulate the success of the VOS model whilst taking into account the presence and evolution of small-scale structure (i.e., wiggles) in the network - to which we are 'blind' in the standard VOS model due to the one-scale approximation. This is achieved by considering that the dynamics of a wiggly Goto-Nambu string can be approximated by that of a smoother (i.e., with no significant structure at scales below $L$) elastic string which obeys the generalized action \cite{CARTERA}
\begin{equation}
S=-\mu_{0}\int\sqrt{-\gamma}\sqrt{1-\gamma^{ab}\phi_{,a}\phi_{,b}}d^{2}\sigma\label{wiggly_S}
\end{equation}
where $\phi$ is a scalar field whose associated current is regarded as a mass current resulting from the propagation of wiggles on the string.
 
Note that $\phi$ is an effective quantity which is related to an undefined renormalization procedure by which structure below some length scale $\ell$ is smoothed. Naturally, $\ell$ should be no greater than the string correlation length, but still large enough for the effective string energy per unit length (and $\phi$) to depend solely on the worldsheet time, at least in regions large enough for an eventual spatial dependence to be negligible in the local equations of motion.

\subsection{\label{quant}Basic properties}

Besides affecting the evolution of the string configuration, the presence of this mass current also changes the way some relevant quantities are defined on the string. 

Given the mesoscopic nature of $\phi$ we can simplify our equations by introducing the dimensionless quantity
\begin{equation}
w=\sqrt{1-\gamma^{ab}\phi_{,a}\phi_{,b}}\label{w}
\end{equation}
in terms of which the local string tension and energy density can be simply written as
\begin{equation}
T=\mu_{0}w\,\quad U=\mu_{0}w^{-1}\label{T_U_w}
\end{equation}

As in the VOS case, the coordinate energy per unit length along the string is given by
\begin{equation}
\epsilon=\sqrt{\frac{\mathbf{x^{\prime}}^2}{1-\mathbf{\dot{x}^{2}}}}\label{epsilon}
\end{equation}
However, there are now two relevant independent energies which can be defined: the total energy in a piece of string
\begin{equation}
E=\mu_{0}a\int{\frac{\epsilon}{w}d\sigma}\label{E}
\end{equation}
and the energy in a Goto-Nambu string with the same configuration as our smoothed elastic string, called the bare energy,
\begin{equation}
E_{0}=\mu_{0}a\int{\epsilon d\sigma}\label{E0}
\end{equation}

Since it is generally assumed that the basic VOS assumptions apply to the smoothed string, it is the bare energy that should be associated with the network correlation length via
\begin{equation}
\rho_{0}=\frac{\mu_{0}}{\xi^{2}}\label{xi}
\end{equation}

Analogously, there are now two natural averaging procedures defined for a generic quantity $Q$ by
\begin{equation}
\left\langle Q\right\rangle =\frac{\int Q\frac{\epsilon}{w}d\sigma}{\int\frac{\epsilon}{w}d\sigma}\label{average}
\end{equation}
and
\begin{equation}
\left\langle Q\right\rangle _{0}=\frac{\int Q\epsilon d\sigma}{\int\epsilon d\sigma}\label{average0}
\end{equation}
the former appearing more naturally in our equations but the latter possibly being more convenient to use in applications when the wiggliness of a string is not well known. Note that, in an infinite string, the two procedures are equivalent if and only if $Q$ is independent of $w$ (i.e., $\left<Qw\right>=\left<Q\right>\left<w\right>$).

Finally, these concepts can be combined in the definition of the renormalized string mass per unit length factor
\begin{equation}
\mu \equiv \frac{E}{E_{0}}\equiv\frac{\xi^{2}}{L^{2}}=\left\langle w\right\rangle ^{-1}=\left\langle w^{-1}\right\rangle _{0}\label{mu}
\end{equation}
which is trivially at least unity ($\mu=1$ corresponding to the Goto-Nambu limit, when there is no small-scale structure) and quantifies the wiggliness of a network.

\subsection{\label{aver}Averaged evolution}

The system of equations which define the model introduced in the previous paper \cite{PAP1} can be found by using the equations of motion obtainable from the action given by Eq.~(\ref{wiggly_S}) together with the following phenomenological terms that model energy loss to loops as well as energy transfer from the bare to the wiggly component due to kink formation by intercommutation
\begin{equation}
\left(\frac{1}{\rho}\frac{d\rho}{dt}\right)_{loops}=-cf\left(\mu\right)\frac{v}{\xi}\label{f}
\end{equation}
\begin{equation}
\left(\frac{1}{\rho_{0}}\frac{d\rho_{0}}{dt}\right)_{loops}=-cf_{0}\left(\mu\right)\frac{v}{\xi}\label{f0}
\end{equation}
\begin{equation}
\left(\frac{1}{\rho_{0}}\frac{d\rho_{0}}{dt}\right)_{wiggles}=-cs\left(\mu\right)\frac{v}{\xi}\label{s}
\end{equation}
where $v\equiv \left\langle\mathbf{\dot{x}}^2\right\rangle$, $c$ is a constant of order unity which corresponds to the loop-chopping parameter of the VOS model, and $f$, $f_{0}$, and $s$ are functions of $\mu$ which are unity (in the case of $f$ and $f_{0}$) and zero (in the case of $s$) if $\mu=1$, lest we not recover the VOS model in the Goto-Nambu limit.

Apart from these energy loss mechanisms, it is important to take into account that varying the renormalization scale $\ell$ is tantamount to redefining what small-scale structure is, and thus must have an effect on the value of $E_{0}$ (as well as $v$ since $w$ is also changed). This can be done by introducing the following scale-drift terms
\begin{equation}
\frac{1}{\mu}\frac{\partial\mu}{\partial\ell}\frac{d\ell}{dt}\sim\frac{d_{m}-1}{\ell}\frac{d\ell}{dt}\label{dm}
\end{equation}
\begin{equation}
\frac{\partial v^{2}}{\partial\ell}\frac{d\ell}{dt}=\frac{1-v^{2}}{1+\left\langle w^{2}\right\rangle}\frac{\partial\left\langle w^{2}\right\rangle}{\partial\ell}\frac{d\ell}{dt}\label{v_dm}
\end{equation}
where $d_{m}\left(\ell\right)$ is the multifractal dimension of a string segment at scale $\ell$ \cite{TAKAYASU}. Note that Eq.~(\ref{dm}) is essentially just a geometric identity whereas Eq.~(\ref{v_dm}) comes from imposing total energy conservation across different scales.

If we further assume uniform wiggliness (i.e., $w$ to be just a function of time) then the system of equations we are looking for is just
\begin{widetext}
\begin{equation}
2\frac{d\xi}{dt}=H\xi\left[2+\left(1+\frac{1}{\mu^{2}}\right)v^{2}\right]+v\left[k\left(1-\frac{1}{\mu^{2}}\right)+c\left(f_{0}+s\right)\right]
+\left[d_{m}\left(\ell\right)-1\right]\frac{\xi}{\ell}\frac{d\ell}{dt}\label{qui_full}
\end{equation}
\begin{equation}
\frac{dv}{dt}=\left(1-v^{2}\right)\left[\frac{k}{\xi\mu^{2}}-Hv\left(1+\frac{1}{\mu^{2}}\right)-\frac{1}{1+\mu^{2}}\frac{\left[d_{m}\left(\ell\right)-1\right]}{v\ell}\frac{d\ell}{dt}\right]\label{v_full}
\end{equation}
\begin{equation}
\frac{1}{\mu}\frac{d\mu}{dt}=\frac{v}{\xi}\left[k\left(1-\frac{1}{\mu^{2}}\right)-c\left(f-f_{0}-s\right)\right]-H\left(1-\frac{1}{\mu^{2}}\right)
+\frac{\left[d_{m}\left(\ell\right)-1\right]}{\ell}\frac{d\ell}{dt}\label{mu_full}
\end{equation}
\end{widetext}
where $H\equiv\dot{a}/a$ is the Hubble parameter and $k$, called the momentum parameter, is defined as
\begin{equation}
k=\frac{\left\langle \left(1-\mathbf{\dot{x}}^{2}\right)\left(\mathbf{\dot{x}\cdot\hat{u}}\right)\right\rangle }{v\left(1-v^{2}\right)} \sim \frac{\left\langle \mathbf{\dot{x}\cdot\hat{u}}\right\rangle}{v}\label{momentum_parameter}
\end{equation}
and in the relevant relativistic regime it can be written as (see \cite{MS3}) 
\begin{equation}
k\left(v\right)=\frac{2\sqrt{2}}{\pi}\frac{1-8v^{6}}{1+8v^{6}}\,. \label{momentum_parameter_ansatz}
\end{equation}

Note that in order for this formalism to be consistent it is already necessary that the uniform wiggliness condition be locally true, even though it can still not be so over cosmological length scales (i.e., $w^{\prime}$ can be very small but non-zero). 

Some interesting considerations can be drawn from the fact that Eqs.~(\ref{dm}--\ref{v_dm}) can be integrated. The former trivially yields
\begin{equation}
\log{[\mu(\ell)]}=\int_0^\ell[d_m(\ell')-1]d\ln{\ell'}\label{solvdm}
\end{equation}
while for the latter, assuming uniform wiggliness and defining the convenient parameter
\begin{equation}
X\equiv\frac{1}{\mu^{2}}\label{X}
\end{equation}
we have
\begin{equation}
v^{2}\left(\ell\right)=1-2\frac{1-v^{2}\left(\ell=0\right)}{1+X\left(\ell\right)} \label{vrms}
\end{equation}
which is an important equation linking a 'microscopic' velocity to wiggliness, and which forces us to face a non-trivial crossroads.

The most natural way to proceed is clearly to keep to the spirit of the VOS model and just interpret the velocity for $\ell=0$ as the RMS velocity that was seen in that model. 
\begin{equation}
v^{2}\left(\ell\right)=1-\frac{2\left(1-v^{2}_{RMS}\right)}{1+X\left(\ell\right)}. \label{vrms1}
\end{equation}
That interpretation, however, necessarily entails an unexpected limitation to the application of the formalism: since this scale-dependent $v^2$ must still be positive, we have to be beyond our domain of applicability whenever $X\left(\ell\right)<1-2v^{2}_{RMS}$. In other words, we should expect our wiggly models to break down in the non-relativistic regime. In particular, this means that our formalism cannot make trustworthy predictions in the tensionless limit. If so, the calculations in this limit in the previous paper worked only because $v$ was artificially fixed at $v=0$ (although the calculations for a fixed $\ell$ should still hold). Even though there is in principle no reason why our formalism should be valid all the time (including in regimes in which the VOS model has not been properly tested) this should at least serve as motivation to entertain a possible alternative.

A perhaps more serious motivation for questioning the validity of Eq.~\ref{vrms1} is related to a certain tension between different types of simulations regarding what this microscopic velocity should be. The RMS velocity measured in expanding universe Goto-Nambu simulations is close to, but slightly below $1/\sqrt{2}$ (highlighting the presence of small-scale wiggles), whereas in Minkowski space Goto-Nambu simulations or field theory simulations the measured velocities are consistent with $1/\sqrt{2}$. This might motivate an even simpler form for the scale dependence of the characteristic velocity,
\begin{equation}
v^{2}(\ell)=\frac{1}{1+\mu^2(\ell)}; \label{vvrms}
\end{equation}
which as we shall see is qualitatively (though not quantitatively) in agreement with numerical simulations if we interpret $v$ as the coherent velocity.

In the end, it seems that which formula is correct is related to whether Goto-Nambu or field theory simulations are more accurate at the relevant scales---see for example the comparison between both types of simulations in \cite{ABELIAN}. Naturally, Goto-Nambu simulations should never be expected to favour Eq.~\ref{vvrms} over Eq.~\ref{vrms1}, but one should keep in mind that ultimately we want to model realistic networks rather than simply fit the output of any type of simulation.

Moreover, there is even no guarantee that either formula has to be correct. In the same way we have already mentioned there is no a priori reason why our formalism should have to be valid in the tensionless limit, there is no reason why it has to be valid down to arbitrarily small scales; especially if we keep in mind this formalism is based on a 'string renormalization' procedure, connecting wiggly and elastic strings, which we do not fully comprehend (especially when it comes to transforming velocity vectors). All we really need in order to use our evolution equations is that it be valid over a range of scales that includes our choice for $\ell$.

Nevertheless, it should be noted that this dilemma can have a non-trivial effect in the complexity of our equations. If Eq.~\ref{vvrms} is true then we can reduce the number of equations in our system since $v$ and $\mu$ are now completely correlated and thus Eqs.~(\ref{v_full}--\ref{mu_full}) cannot be independent. This realization allows us to relate the loop-chopping terms to the momentum parameter and the Hubble parameter via
\begin{equation}
\frac{v}{\xi} \left[2k - c \left( f - f_{0} - s \right)\right] = 2H \label{cons_muv}
\end{equation}
which in particular implies, since $k\left(v=\frac{1}{\sqrt{2}}\right)=0$, that
\begin{equation}
\xi\left(\ell=0\right)=-c\frac{f\left(1\right)-f_{0}\left(1\right)-s\left(1\right)}{2\sqrt{2}H} \label{cons_cor}
\end{equation}
and the numerator, usually assumed to be null in this limit, now has to be non-zero. This is not wholly unexpected since the null case corresponds to an attempt to recover the VOS model exactly as $\ell$ goes to zero, which this approach must necessarily contradict.

Finally, note that Eqs.~(\ref{vrms}--\ref{vvrms}) are all very useful tools since they provide us with a way to test whether a scale-dependent velocity is the characteristic velocity in our model (independently of the multifractal dimension), which may further our physical understanding of this formalism. Nonetheless, most of the following calculations will only assume Eq.~\ref{vrms} simply  because most simulations available to us are Goto-Nambu and using Eq.~\ref{vvrms} would require knowing more about energy-loss mechanisms (i.e., more freedom  in parametrizing $f$, $f_{0}$, and $s$). Regardless, it would be straightforward to carry out the analogous calculations, which would actually be simpler to solve, as they would typically involve systems of two equations instead of three, with Eq.~\ref{cons_muv} working as a consistency relation among the parameters of the model.

\section{\label{sca}The Scaling Regime}

The prediction of an attractor scaling regime when the scale factor is a power law (as in Eq.~\ref{a_power}) is one of the main predictions of the VOS model which is in quantitative agreement with numerical simulations. This regime is characterized by a constant velocity and a characteristic length proportional to time (or, equivalently, to the cosmological horizon length). Specifically, the VOS model predicts \cite{MS3}
\begin{equation}
\left(\frac{L}{t}\right)^{2}\equiv\gamma^{2}=\frac{k\left(k+c\right)}{4\lambda\left(1-\lambda\right)}\label{GN_gamma}
\end{equation}
\begin{equation}
v^{2}=\frac{k\left(1-\lambda\right)}{\lambda\left(k+c\right)}\label{GN_v}
\end{equation}
and since this result is confirmed by Abelian-Higgs simulations (for $c=0.23$) our corresponding prediction should not significantly deviate from this.

An important open question in cosmic string evolution is whether the small-scale component also scales, i.e., whether we should also expect $\mu$ to evolve towards a constant value. Despite current simulations not answering this question definitely \cite{FRAC}, they suggest that such a small-scale scaling is reached at least in a matter era (when $\lambda=2/3$). In the radiation era simulations show a more complex behavior, which could reflect the fact that the approach to scaling is slower in this case (since there is less Hubble damping) or could be due to the existence of more than one scaling solution.

\subsection{Finding wiggly scaling\label{findscale}}

Scaling solutions can be straightforwardly sought by making the appropriate substitutions on the left-hand side of Eqs.~(\ref{qui_full}--\ref{mu_full}) and assuming that $\ell$ is also scaling. At this point we need to specify a specific behavior for the fractal dimension $d_m$ as a function of the other parameters. (A mathematically simpler but physically less realistic alternative would be to consider it a constant phenomenological parameter at the scale $\xi$ that we'll be interested in.) This turns out to be a more subtle question than it may appear, and a full derivation is left for subsequent work, but we can nevertheless provide an approximate derivation here.

It is obvious that the fractal dimension will be scale-dependent, ranging from $d_m=1$ on very small scales to $d_m=2$ (Brownian) on super-horizon scales, and interpolating between the two limits on scales around the correlation length. Such a behavior has been explicitly shown to occur in Goto-Nambu simulations \cite{FRAC}. We can therefore construct a fairly generic phenomenological function that reproduces this behavior
\begin{equation}
d_m(\ell)=2-\left[1+B\left(\frac{\ell}{\xi}\right)^b\right]^{-1} \,. \label{fractal1}
\end{equation}
This allows freedom both in the characteristic scale at which the transition occurs and in how fast it occurs as one changes scale. Now, the fractal dimension and $\mu$
are related by Eq.~\ref{solvdm} and in this case this yields
\begin{equation}
\mu(\ell)=\left[1+B\left(\frac{\ell}{\xi}\right)^b\right]^{1/b} \,. \label{fractal3}
\end{equation}
By simple substitution we can now remove the $\ell$ dependence and obtain an explicit relation between $d_m$ and $\mu$
\begin{equation}
d_m(\mu)=2-\frac{1}{\mu^b}\,. \label{fractal4}
\end{equation}
Notice that this depends only on the parameter $b$, not on $B$.

All that remains to be done is to fix the free parameter $b$. Comparing to expanding universe numerical simulations \cite{FRAC} we find that $b=2$ provides a fairly reasonable approximation. Thus in what follows we will use
\begin{equation}
d_{m}=2-\frac{1}{\mu^{2}}\label{dm_mu}\,.
\end{equation}
Note that combining this with Eq.~\ref{vrms1} we can also write
\begin{equation}
v^{2}(\ell)=1-\frac{2\left(1-v^{2}_{RMS}\right)}{3-d_m(\ell)}=\frac{1-d_m(\ell)+2v^{2}_{RMS}}{3-d_m(\ell)}\,, \label{vrms2}
\end{equation}
or equivalently
\begin{equation}
d_m(\ell)=3-\frac{2\left(1-v^{2}_{RMS}\right)}{1-v^2(\ell)}=\frac{1+2v^{2}_{RMS}-3v^2(\ell)}{1-v^2(\ell)}\,; \label{vrms3}
\end{equation}
naturally the analogous expressions for the ansatz of Eq.~\ref{vvrms} ensue by taking the particular case $v_{RMS}=1/\sqrt{2}$.

With these assumptions we can now reduce our problem to solving the algebraic system
\begin{equation}
v^{2}=\frac{\left[4X^{2}-2\lambda X\left(1+X\right)\right]\left(k/c\right)-X(1-X)\left(f_{0}+s\right)}{\lambda\left(1+X\right)^{2}\left[\left(k/c\right)+f_{0}+s\right]}\label{full_scaling_system1}
\end{equation}
\begin{equation}
\gamma_{\xi}=v\frac{k\left(1-X\right)+c\left(f_{0}+s\right)}{1+X-\lambda\left[2+\left(1+X\right)v^{2}\right]}\label{full_scaling_system2}
\end{equation}
\begin{equation}
\frac{v}{\gamma_{\xi}}\left[k\left(1-X\right)-c\left(f-f_{0}-s\right)\right]+\left(1-\lambda\right)\left(1-X\right)=0\label{full_scaling_system3}
\end{equation}
which interestingly has at most two solutions with the same fixed value of $X\neq 1$ (assuming that the shape of the energy loss functions is fixed). In other words, for any given $X$ there are at most two values of $c$ such that there is a scaling solution with that constant value of $X$; in what follows we will denote these by $c_{X}$. These solutions, if they exist, can be found by the following algorithm: first just compute
\begin{equation}
v_{X}^{2}=\frac{\left[4X^{2}-2\lambda X\left(1+X\right)\right]\varphi_{X}-X(1-X)\left(f_{0}+s\right)}{\lambda\left(1+X\right)^{2}\left[\varphi_{X}+f_{0}+s\right]}\label{eq:scaling_cons_V_func_X}
\end{equation}
where $\varphi_{X}$ is a real solution of the quadratic equation
\begin{equation}
A\varphi_{X}^{2}+B\varphi_{X}+C=0\label{eq:varphi_X}
\end{equation}
whose coefficients are
\begin{widetext}
\begin{equation}
A=\left(1-\lambda\right)\left(1-X\right)\left(1-X^{2}\right) 
-\left(1-X\right)\left[4X^{2}-2\lambda\left(1+X\right)X\right]+\left(1-X^{2}\right)\left[1+X-2\lambda\right] \label{eq:scalingA}
\end{equation}
\begin{eqnarray}
B &=& \left(1-\lambda\right)\left(1-X^{2}\right)\left(2-X\right)\left(f_{0}+s\right)+\left(f-f_{0}-s\right)\left(4X^{2}-2\lambda\left(1+X\right)X\right)\\ \nonumber
 &+& \left(f_{0}+s\right)X\left(1-X\right)^{2} +\left[\left(f_{0}+s\right)\left(1-X\right)-f+f_{0}+s\right]\left[\left(1+X\right)^{2}-2\lambda\left(1+X\right)\right]\label{eq:scalingB}
\end{eqnarray}
\begin{equation}
C=\left(f_{0}+s\right)^{2}\left(1-\lambda\right)\left(1-X^{2}\right) 
-\left(f_{0}+s\right)\left(f-f_{0}-s\right)\left[X\left(1-X\right)+\left(1+X\right)^{2}-2\lambda\left(1+X\right)\right]\label{eq:scalingC}
\end{equation}
\end{widetext}
(of course, if there are no real solutions to Eq.~\ref{eq:varphi_X} that
just means that scaling is impossible for that $X$), then compute
$k\left(v_{X}\right)$ using Eq.~\ref{momentum_parameter_ansatz} and
the $c_{X}$ we are after is simply 
\begin{equation}
c_{X}=\frac{k\left(v_{X}\right)}{\varphi_{X}}\label{eq:scaling_c}
\end{equation}
if it is positive and less than $1$ - otherwise there is no scaling. Obviously, there is also no scaling if the velocity $v$ and the correlation coefficient $\gamma_{\xi}$ calculated in this way have non-physical values.

Interestingly, one can see by setting $X=1$ that the VOS solutions are also solutions of our model provided that $f_{0}\left(X=1\right)=f\left(X=1\right)=1$ and $s\left(X=1\right)=0$. That is by no means unexpected, since when building this model we required that the VOS equations be recovered whenever $X=1$, $f_{0}=f=1$, and $s=0$. This is not to be regarded as a problem since $s\left(X=1\right)=0$ is an approximation which is to some extent motivated by the success of the VOS predictions. In a way, we are just saying that $s\left(X=1\right)$ gives a contribution which is much weaker than those of competing energy loss mechanisms.

\subsection{Wiggly scaling stability\label{stable}}

Ultimately, the feature that made scaling regimes in the VOS model interesting was their attractor nature - which, in particular, enables us to use them to calibrate the loop-chopping efficiency $c$ by comparison with simulations. Therefore, a study of the stability of the non-trivial (here meaning those with $X\neq 1$) scaling solutions found above is needed.

With this in mind, it is straightforward to linearize our equations around these solutions
\begin{equation}
\left[\begin{array}{c}
\gamma_{\xi,}\\
v_{\,}\\
X_{\,}
\end{array}\right]\sim\left[\begin{array}{c}
\gamma_{ s}\\
v_{s}\\
X_{s}
\end{array}\right]+\left[\begin{array}{c}
\overline{\gamma_{\xi}}\\
\overline{v}\\
\bar{X}
\end{array}\right]\label{linearize}
\end{equation}
and write them in matrix form
\begin{equation}
t\frac{d}{dt}\left[\begin{array}{c}
\overline{\gamma_{\xi}}\\
\overline{v}\\
\bar{X}
\end{array}\right]=\left[\begin{array}{ccc}
\\
&M_{j}^{i}\\
\\
\end{array}\right]\left[\begin{array}{c}
\overline{\gamma_{\xi}}\\
\overline{v}\\
\bar{X}
\end{array}\right]\label{stability_M}
\end{equation}
where $\gamma_{s}$, $v_{s}$, and $X_{s}$ are the scaling values of $\gamma_{\xi}$, $v$, and $X$, respectively. The components of $M$ can be shown to be
\begin{widetext}
\begin{eqnarray}
M_{1}^{1}&=&-1+\lambda\left(\frac{2}{1+X_{s}}+v_{s}^{2}\right)\\
M^{1}_{2}&=&2\lambda\gamma_{s}v_{s}+\frac{B_{s}+\left(k_{s}+v_{s}k_{\star}\right)\left(1-X_{s}\right)}{1+X_{s}} \\
M^{1}_{3}&=&\frac{v_{s}\left(\lambda\gamma_{s}v_{s}-k_{s}+B_{\star}\right)}{1+X_{s}} \\
M^{2}_{1}&=&\left(1-v_{s}^{2}\right)\left(-\frac{\lambda v_{s}}{\gamma_{s}}\left[1+X_{s}\right]-\frac{X_{s}\left[1-X_{s}\right]}{\gamma_{s}v_{s}\left[1+X_{s}\right]}\left(1+M_{1}^{1}\right)\right)\\
M^{2}_{2}&=&\left(1-v_{s}^{2}\right)\left(\frac{X_{s}k_{s}}{\gamma_{s}v_{s}}+\frac{k_{\star}}{\gamma_{s}}-2\lambda\left[1+X_{s}\right]-\frac{X_{s}\left[1-X_{s}\right]}{\gamma_{s}v_{s}\left[1+X_{s}\right]}M_{2}^{1}\right)-\frac{2k_{s}v_{s}X_{s}}{\gamma_{s}}+2\lambda v_{s}^{2}\left(1+X_{s}\right)+\frac{2X_{s}\left(\-X_{s}\right)}{1+X_{s}} \\
M^{2}_{3}&=&\left(1-v_{s}^{2}\right)\left(\frac{k_{s}\left[1+2X_{s}\right]}{\gamma_{s}\left[1+X_{s}\right]}-2\lambda v_{s}-\frac{\left(1-2X_{s}\right)}{v_{s}\left(1+X_{s}\right)}-\frac{X_{s}\left[1-X_{s}\right]}{\gamma_{s}v_{s}\left[1+X_{s}\right]}M_{3}^{1}\right) \\
M^{3}_{1}&=&\frac{2\lambda X_{s}\left(1-X_{s}\right)}{\gamma_{s}}-\frac{2X_{s}\left(1-X_{s}\right)}{\gamma_{s}}\left[1+M_{1}^{1}\right] \\
M^{3}_{2}&=&-\frac{2X_{s}\left[\left(1-X_{s}\right)v_{s}k_{\star}+\left(k_{s}\left[1-X_{s}\right]-M_{s}\right)\right]}{\gamma_{s}}-\frac{2X_{s}\left(1-X_{s}\right)}{\gamma_{s}}M_{2}^{1} \\
M^{3}_{3}&=&2\frac{X_{s}}{\gamma_{s}}v_{s}\left(k_{s}+M_{\star}\right)+2X_{s}\left(1-\lambda\right)-\frac{2X_{s}\left(1-X_{s}\right)}{\gamma_{s}}M_{3}^{1} \label{M_components}
\end{eqnarray}
\end{widetext}
where
\begin{equation}
k\equiv k_{s}+k_{\star}\bar{v}\,,
\end{equation}
meaning that
\begin{equation}
k_{\star}=-k_{s}\frac{96v_{s}^{5}}{1-64v_{s}^{12}}\,,
\end{equation}
\begin{equation}
c\left(f_{0}+s\right)\equiv B_{s}+B_{\star}\bar{X}\,,
\end{equation}
and
\begin{equation}
c\left(f-f_{0}-s\right)\equiv M_{s}+M_{\star}\bar{X}\,.
\end{equation}
In writing these formulas for the components of $M$ one also has to assume the natural relation for the mesoscopic scale $\ell$
\begin{equation}
\ell\propto\xi\label{ell_prop}
\end{equation}
which is logical given that $\xi$ is the most important scale governing loop production (not to mention that it scales), but similar expressions for the components of $M$ could be found by assuming any alternative of similar form.

If $u_{k}$ (with $k=1,2,3$) are the eigenvectors of $M^{i}_{j}$ with eigenvalues $\alpha_{k}$ then it is easily seen that $u_{k}\propto t^{\alpha_{k}}$ and, in particular, a scaling solution is stable in this linearized limit if and only if the real parts of all three eigenvalues of $M^{i}_{j}$ are negative. Therefore, whether and how fast our three independent variables approach their scaling values is completely determined by the values of the three eigenvalues (and respective eigenvectors) of the matrix $M$. 

\begin{figure*}[!h]
\includegraphics[scale=0.3]{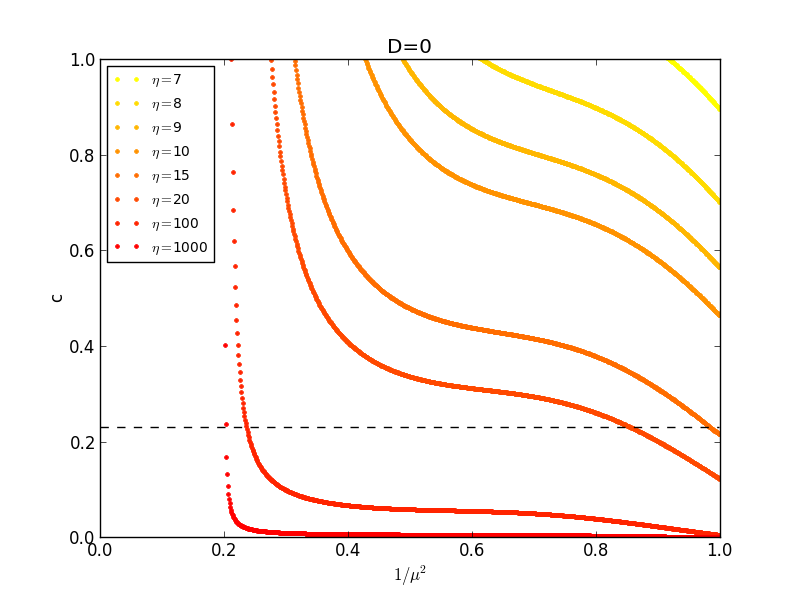}
\includegraphics[scale=0.3]{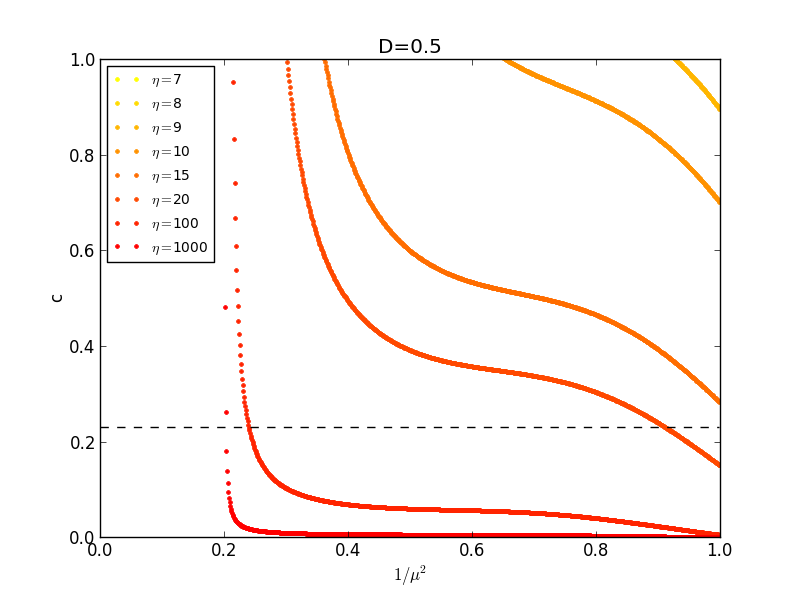}
\includegraphics[scale=0.3]{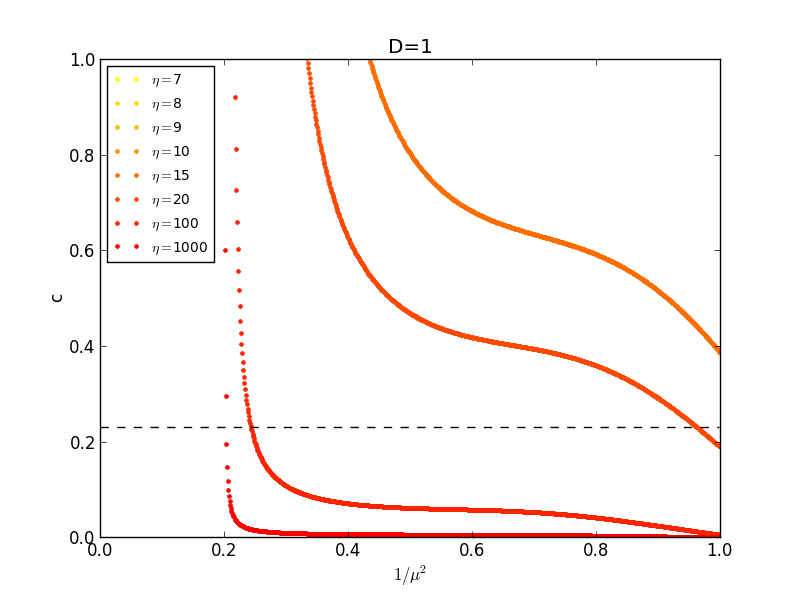}
\includegraphics[scale=0.3]{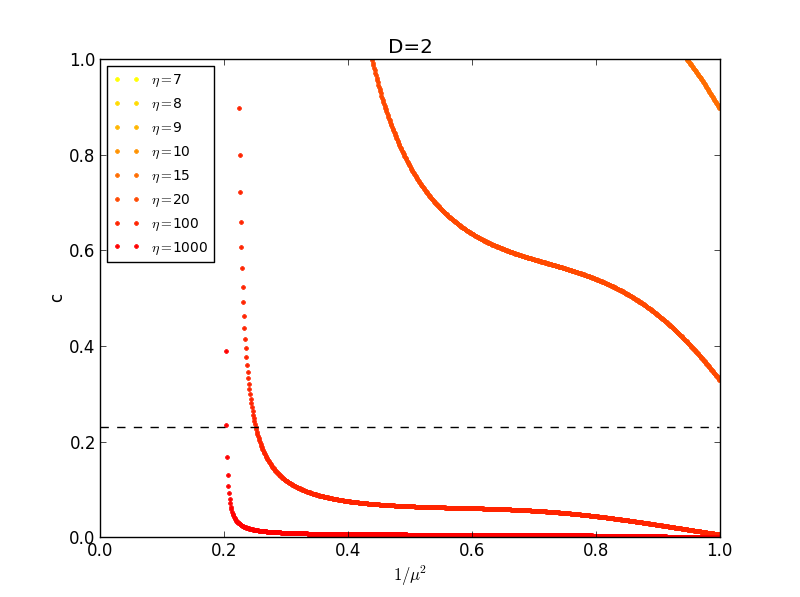}
\caption{\label{fig_matter} Values of the loop-chopping parameter $c$ for which there can be non-trivial scaling, as a function of wiggliness during scaling and for different values of $D$, calculated in the matter era. The dashed line is $c=0.23$, the best fit for the VOS model (the best fit for our model does not have to be the same, but we expect it to be close). We only show the physically meaningful values that stem from Eq.~\protect\ref{eq:varphi_X}---the complementary solution would lead to non-physical (negative) values of $c$.}
\end{figure*}

\begin{figure*}[!h]
\includegraphics[scale=0.3]{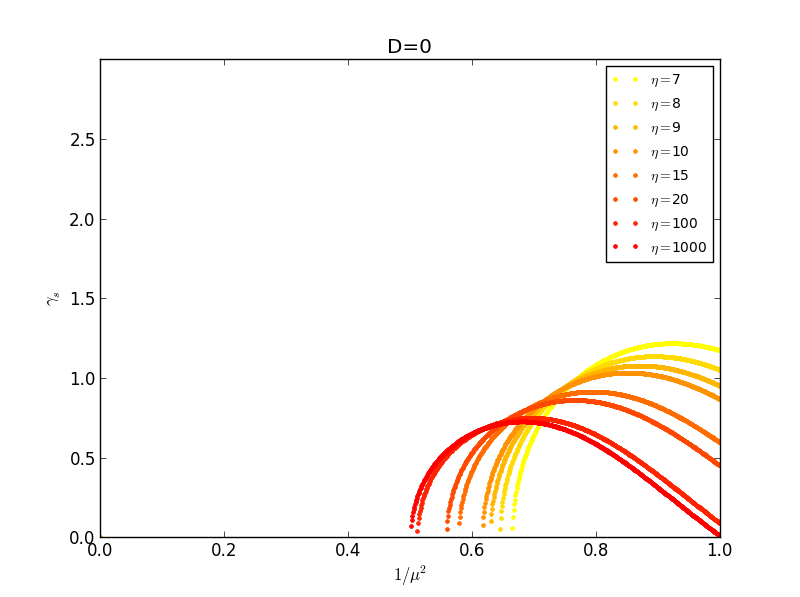}
\includegraphics[scale=0.3]{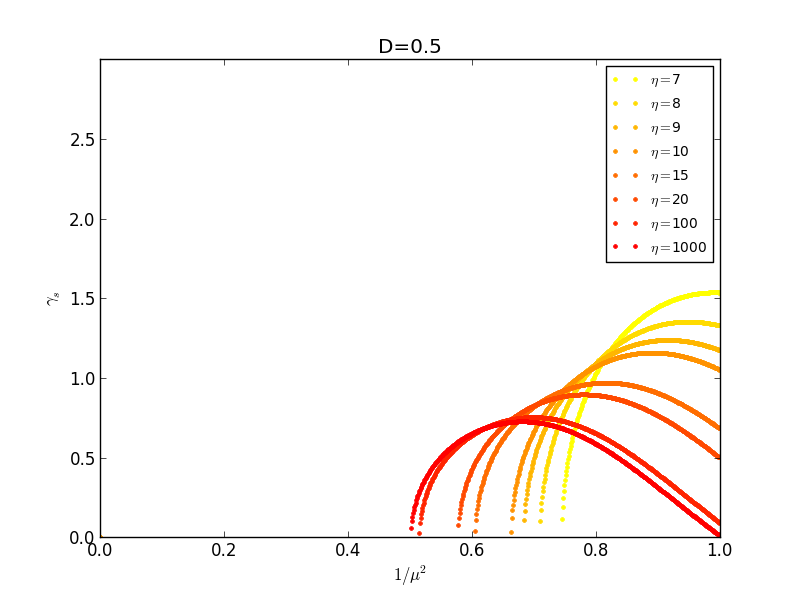}
\includegraphics[scale=0.3]{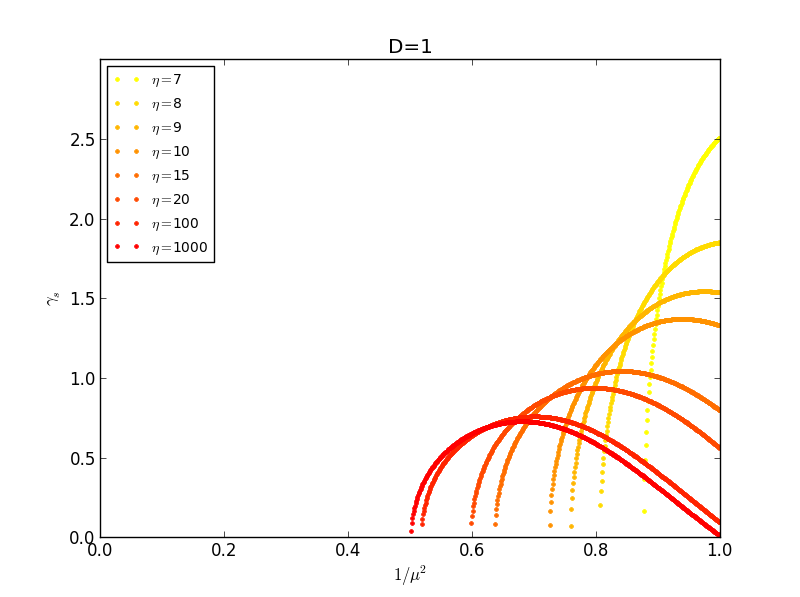}
\includegraphics[scale=0.3]{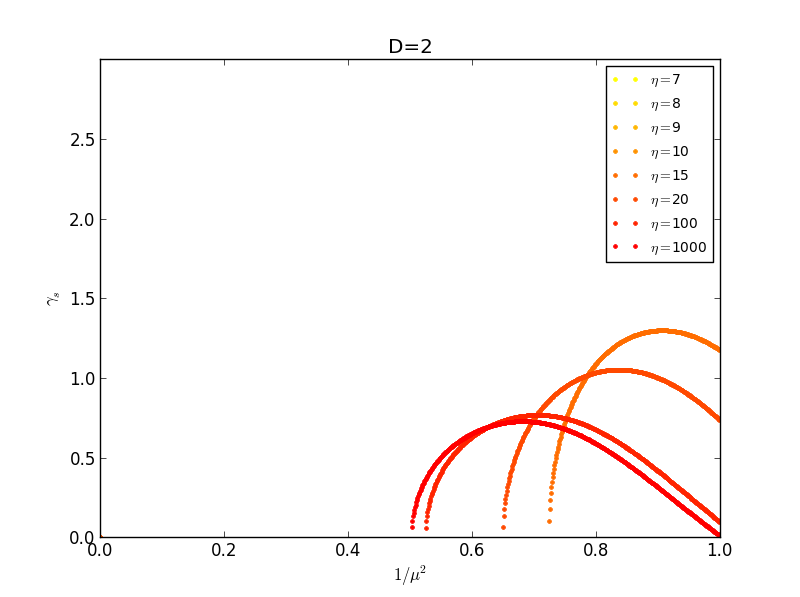}
\caption{\label{fig_matterg} Values of the correlation scaling parameter for which there can be non-trivial scaling, as a function of wiggliness during scaling and for different values of $D$ in the matter era. As before, only physically meaningful values are shown(in this case, $0<\gamma_{s}<\frac{1}{1-\lambda}$).}
\end{figure*}

\begin{figure*}[!h]
\includegraphics[scale=0.3]{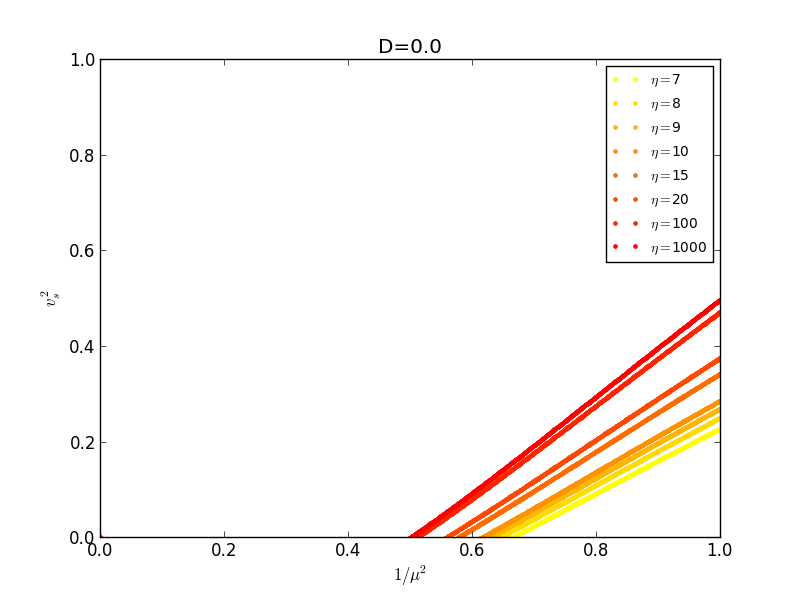}
\includegraphics[scale=0.3]{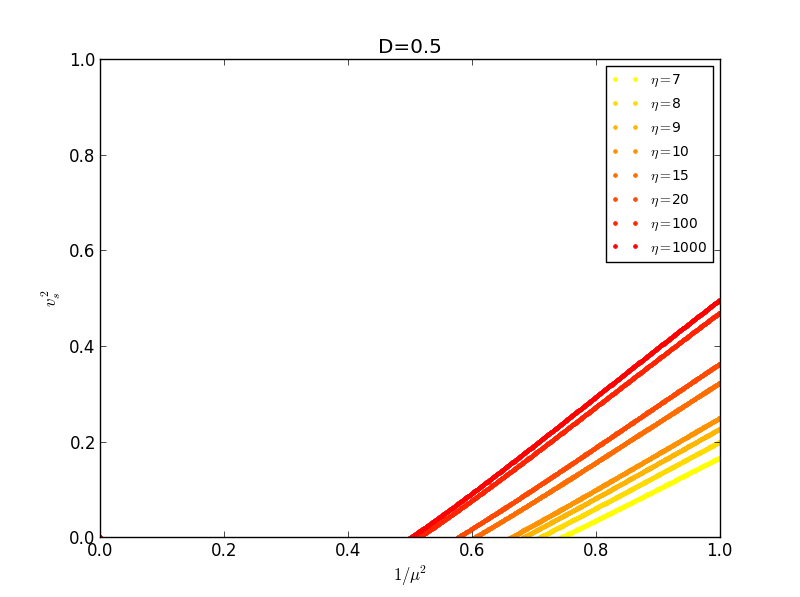}
\includegraphics[scale=0.3]{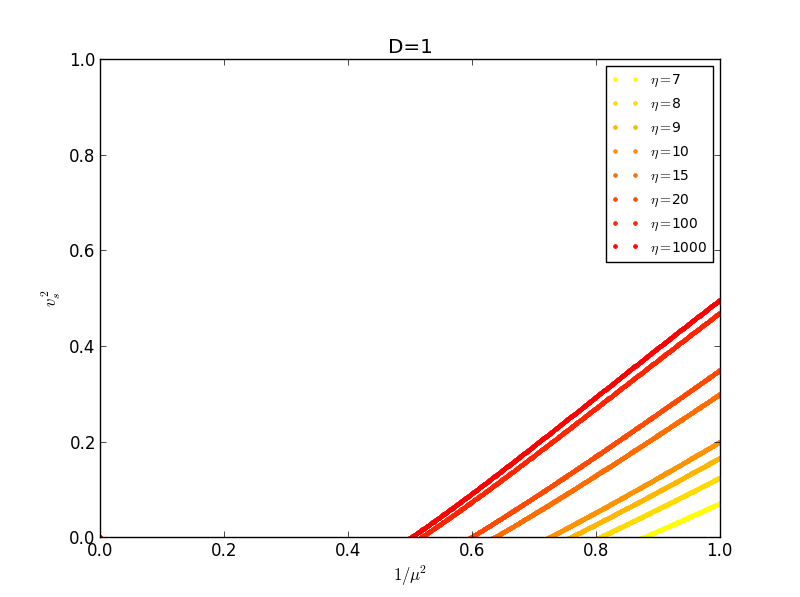}
\includegraphics[scale=0.3]{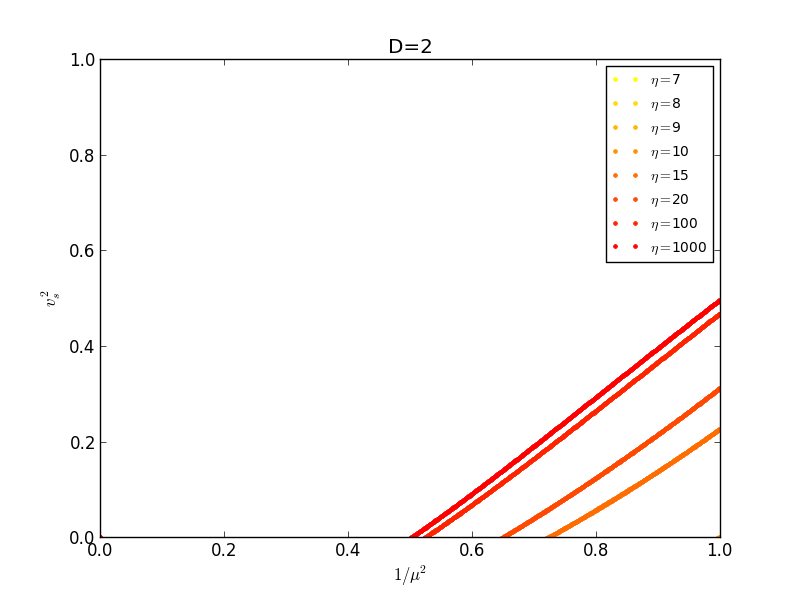}
\caption{\label{fig_matterv} Values of the velocity for which there can be non-trivial scaling, as a function of wiggliness during scaling and for different values of $D$ in the matter era. As before, only physically meaningful values are shown(in this case, $0<v^{2}_{s} <1$).}
\end{figure*}

\begin{figure*}[!h]
\includegraphics[scale=0.3]{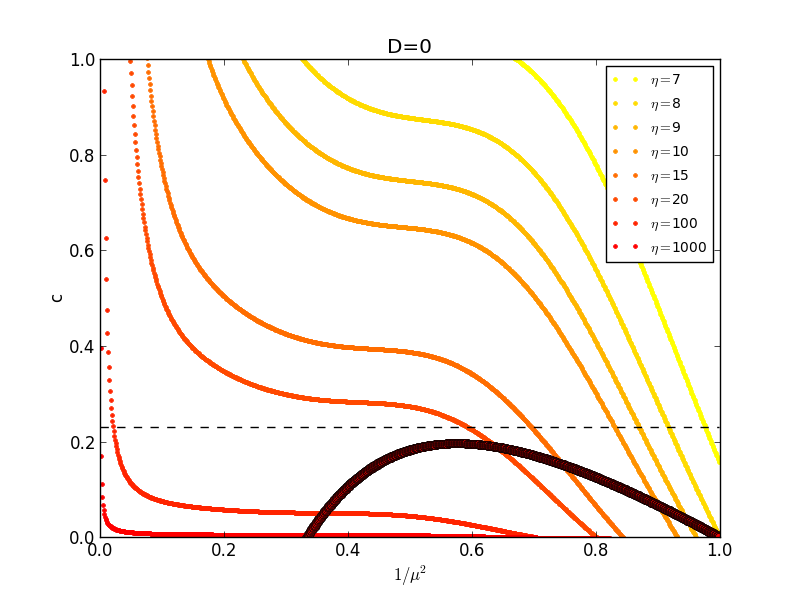}
\includegraphics[scale=0.3]{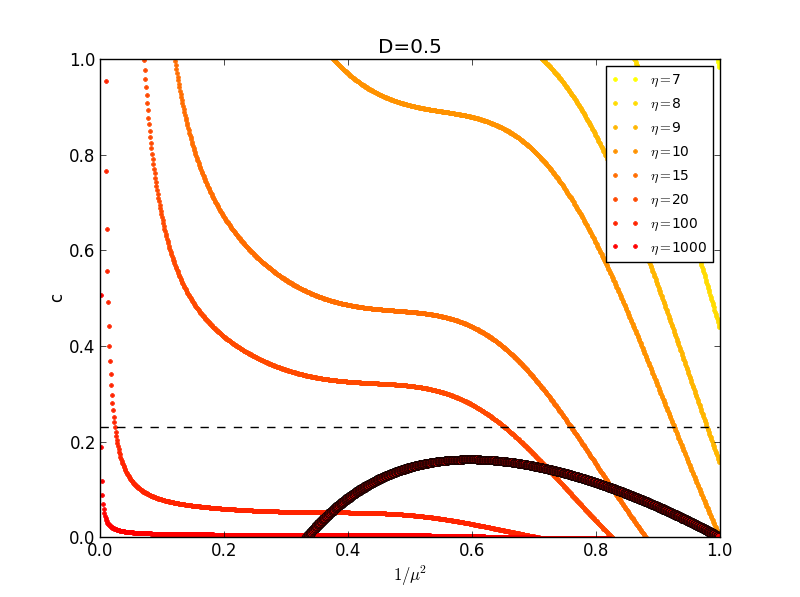}
\includegraphics[scale=0.3]{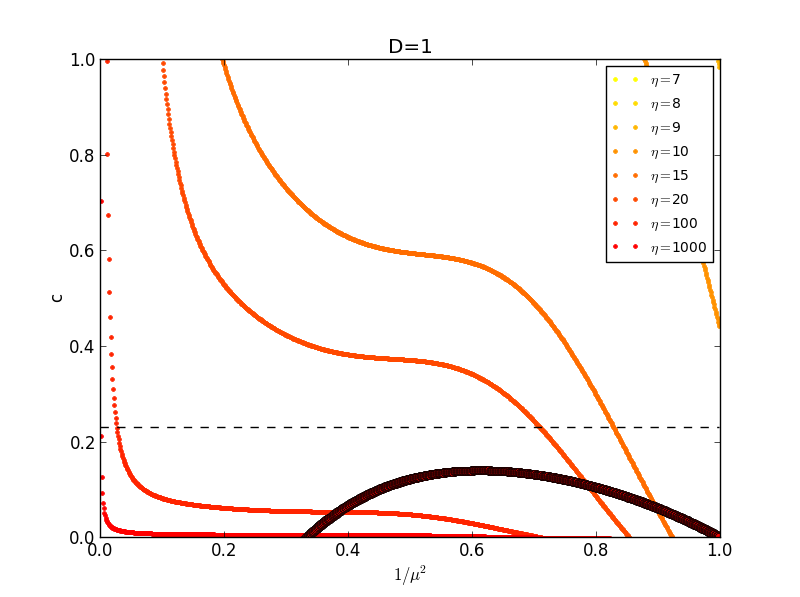}
\includegraphics[scale=0.3]{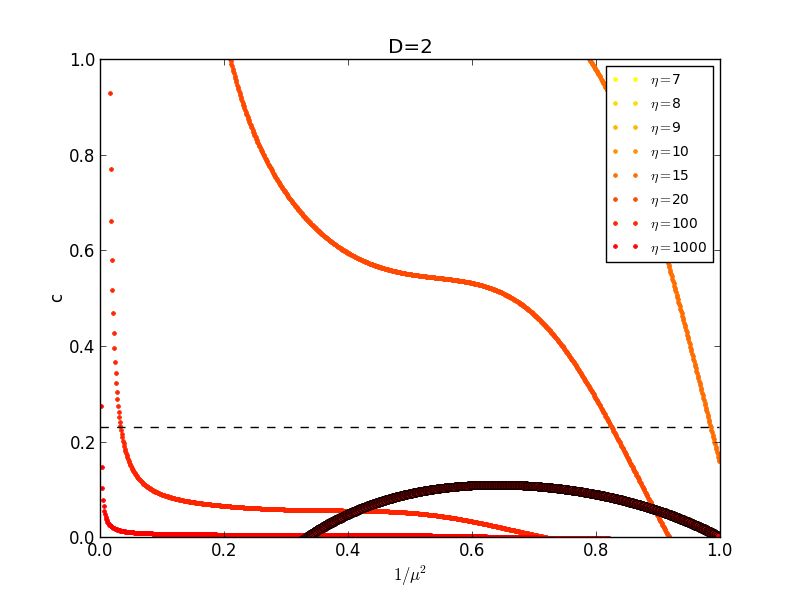}
\caption{\label{fig_rad} Values of the loop-chopping parameter $c$ for which there can be non-trivial scaling, as a function of wiggliness during scaling and for different values of $D$, calculated in the radiation era. The dashed line is $c=0.23$, the best fit for the VOS model (the best fit for our model does not have to be the same, but we expect it to be close). The darker line is there essentially because points of all colors are being plotted on top of each other. Notice that we are only showing the physically meaningful values (in this case, $0<c<1$).}
\end{figure*}

\begin{figure*}[!h]
\includegraphics[scale=0.3]{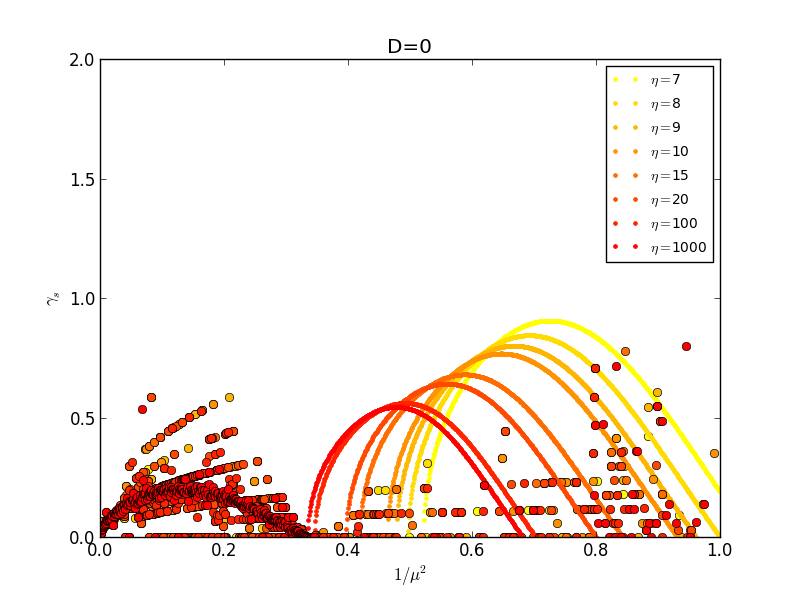}
\includegraphics[scale=0.3]{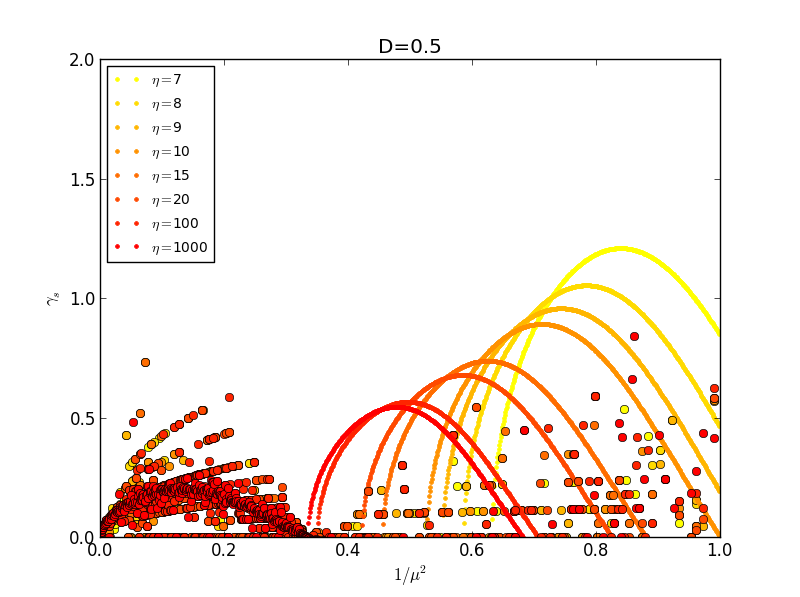}
\includegraphics[scale=0.3]{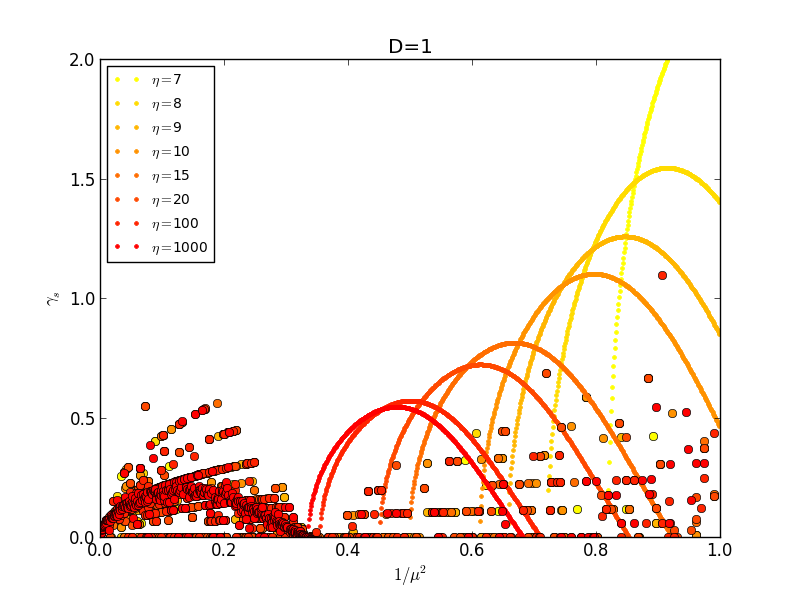}
\includegraphics[scale=0.3]{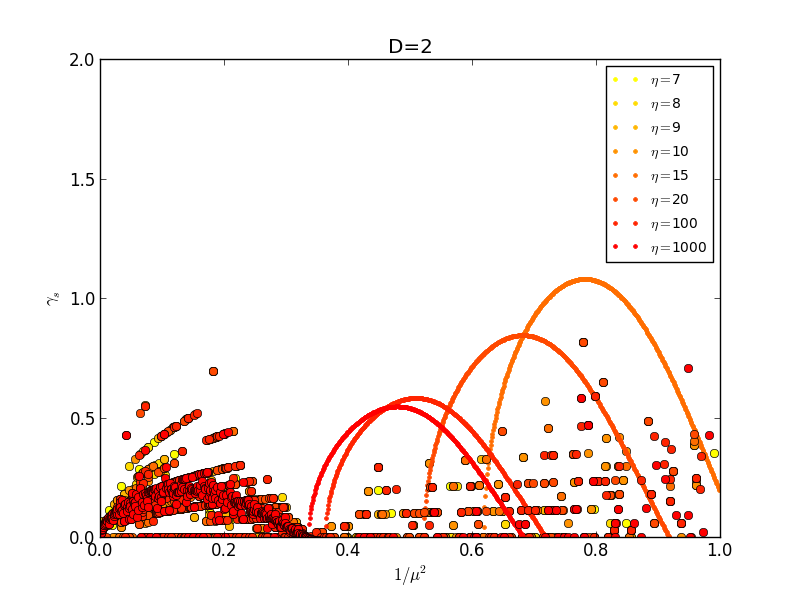}
\caption{\label{fig_radg} Values of the correlation scaling parameter for which there can be non-trivial scaling, as a function of wiggliness during scaling and for different values of $D$ in the radiation era. As before, only physically meaningful values are shown(in this case, $0<\gamma_{s}<\frac{1}{1-\lambda}$). Note that these graphs are fairly contaminated by "noise" generated by computational errors.}
\end{figure*}

\begin{figure*}[!h]
\includegraphics[scale=0.3]{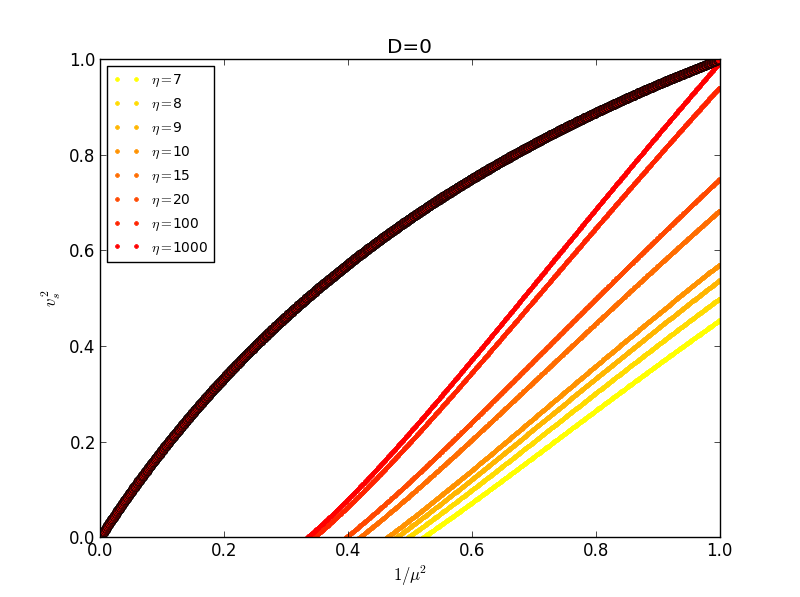}
\includegraphics[scale=0.3]{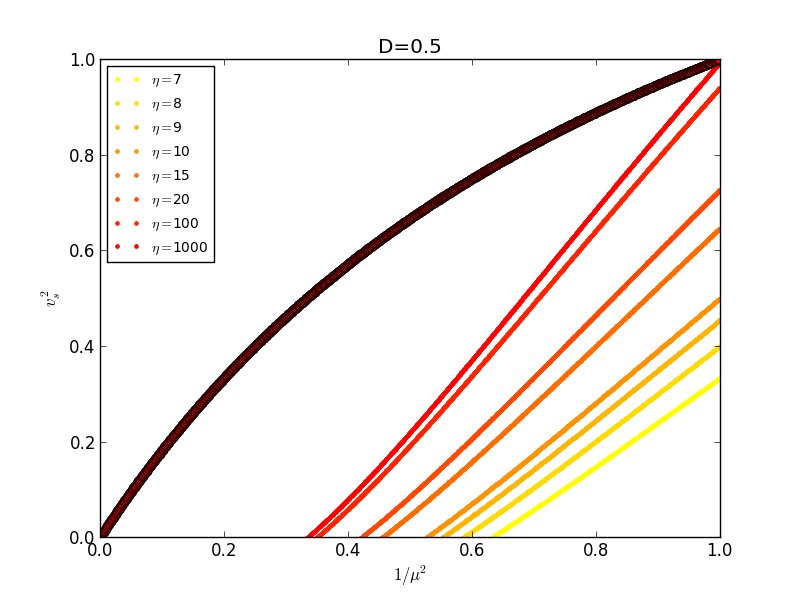}
\includegraphics[scale=0.3]{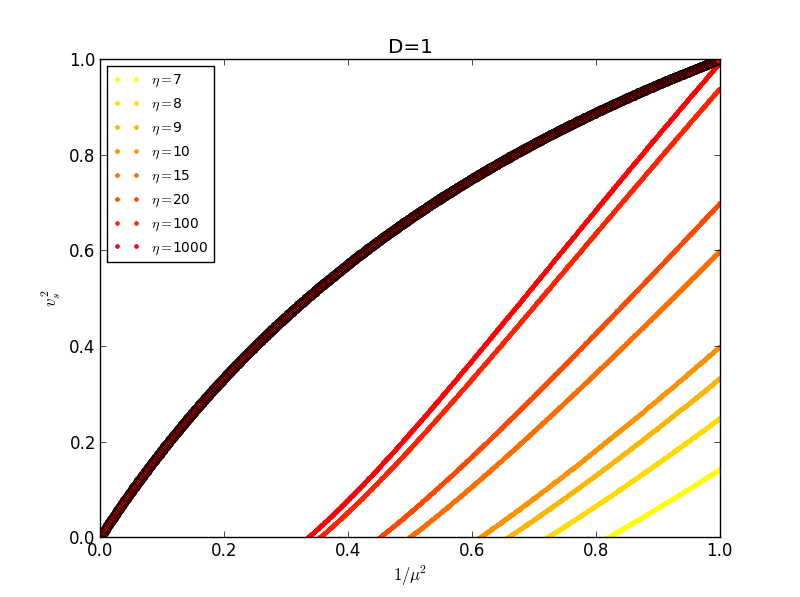}
\includegraphics[scale=0.3]{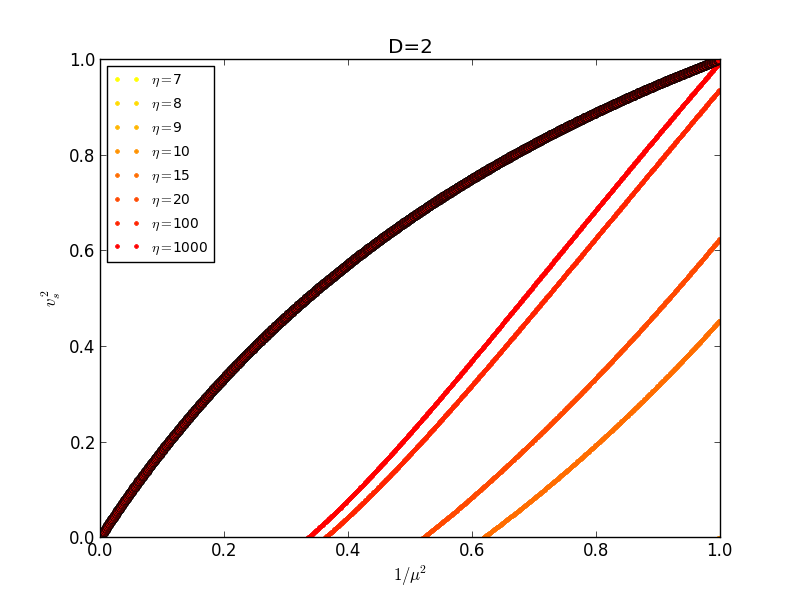}
\caption{\label{fig_radv} Values of the velocity for which there can be non-trivial scaling, as a function of wiggliness during scaling and for different values of $D$ in the radiation era.  The darker line is there essentially because points of all colors are being plotted on top of each other. As before, only physically meaningful values are shown (in this case, $0<v^{2}_{s} <1$).}
\end{figure*}

\subsection{Exploring Scaling\label{scasol}}

Let us now illustrate the procedure described above, starting by introducing a particular ansatz for the energy loss terms. As has been noted, the dependence of $f$ and $f_{0}$ on $\mu$ can in principle be investigated using high-resolution network simulations. In the absence of such information, however, when forced to consider a specific type of dependence, we shall resort to a more ad-hoc argument.

Recall that when the loop-chopping parameter $c$ is introduced in one-scale-type models it is usually as a result of the appearance of a loop-production function, $g$, which only depends on the ratio between the size of loops being produced and the correlation length of the network. This is typically defined \cite{VSH} so that
\begin{equation}
\left.\frac{d\rho_{0}}{dt}\right|_{loops}=-\frac{\mu_{0}v}{\xi^{3}}\intop_{0}^{\infty}g\left(l/\xi\right)\frac{dl}{\xi}\equiv-cv\frac{\rho_{0}}{\xi}\label{OS_loop_g}\,.
\end{equation}
Since we generally assume that the bare string is one for which the VOS assumptions apply, it makes sense to not change this relation and simply use
\begin{equation}
f_{0}=1\label{f0_ansatz}\,.
\end{equation}

Bearing in mind that deviations in the total energy lost to loops should be due to a second loop-production mechanism operating on a scale significantly smaller than the correlation length, it makes sense to expect that $f>1$. Furthermore we conjecture that, in the context of this formalism, the typical length of these smaller loops can be related to a combination of $L$ and $\xi$ that vanishes in the Goto-Nambu limit, when $L=\xi$. Clearly, the simplest such scale is just $\xi_{\star}=\xi-L$. We are then justified to write
\begin{equation}
\left.\frac{d\rho}{dt}\right|_{loops}=-\frac{\mu_{0}\mu v}{\xi^{3}}\intop_{0}^{\infty}g\left(l/\xi\right)\frac{dl}{\xi}-\frac{\mu_{0}\mu v}{\xi^{3}}\intop_{0}^{\infty}g_{\star}\left(l/\xi_{\star}\right)\frac{dl}{\xi}\label{wiggly_loop_g}
\end{equation}
 which corresponds to 
\begin{equation}
f\left(\mu\right)=1+\eta\left(1-\frac{1}{\sqrt{\mu}}\right)\label{f_ansatz}
\end{equation}
where we have defined $\eta=c^{-1}\int{g_{\star}\left(x\right)dx}$, which is a positive parameter quantifying how much energy is lost to small-scale loops. For the sake of simplicity, let us further assume that $s$ can be approximated by
\begin{equation}
s\left(\mu\right)\simeq D\left(1-\frac{1}{\mu^{2}}\right)\label{D_def}
\end{equation}
which we expect to be the case as long as $\mu$ is not too large.

To begin with, let us look for non-trivial scaling solutions without worrying about stability; we address the latter issue in the following section. We start by focusing on the matter era ( $\lambda=2/3$), which is when simulations suggest that it is the easiest to achieve scaling \cite{FRAC}. 

Applying the procedure described in subsection \ref{findscale} to find $c_{X}$, we get the results summarized in Fig. \ref{fig_matter}. In accordance with our simplistic interpretation of $\eta$ is the observation that increasing $\eta$ leads to a decrease in the $c_{X}$ necessary to maintain scaling with a fixed wiggliness value (essentially, since more energy is lost per collision, we need not be so efficient at colliding). More counterintuitive is the realization that an increase in $\eta$ for a fixed $c$ leads to a higher scaling wiggliness - one would naively expect the opposite behavior, that more small-scale energy loss led to a lower scaling wiggliness.

\begin{figure*}
\includegraphics[scale=0.3]{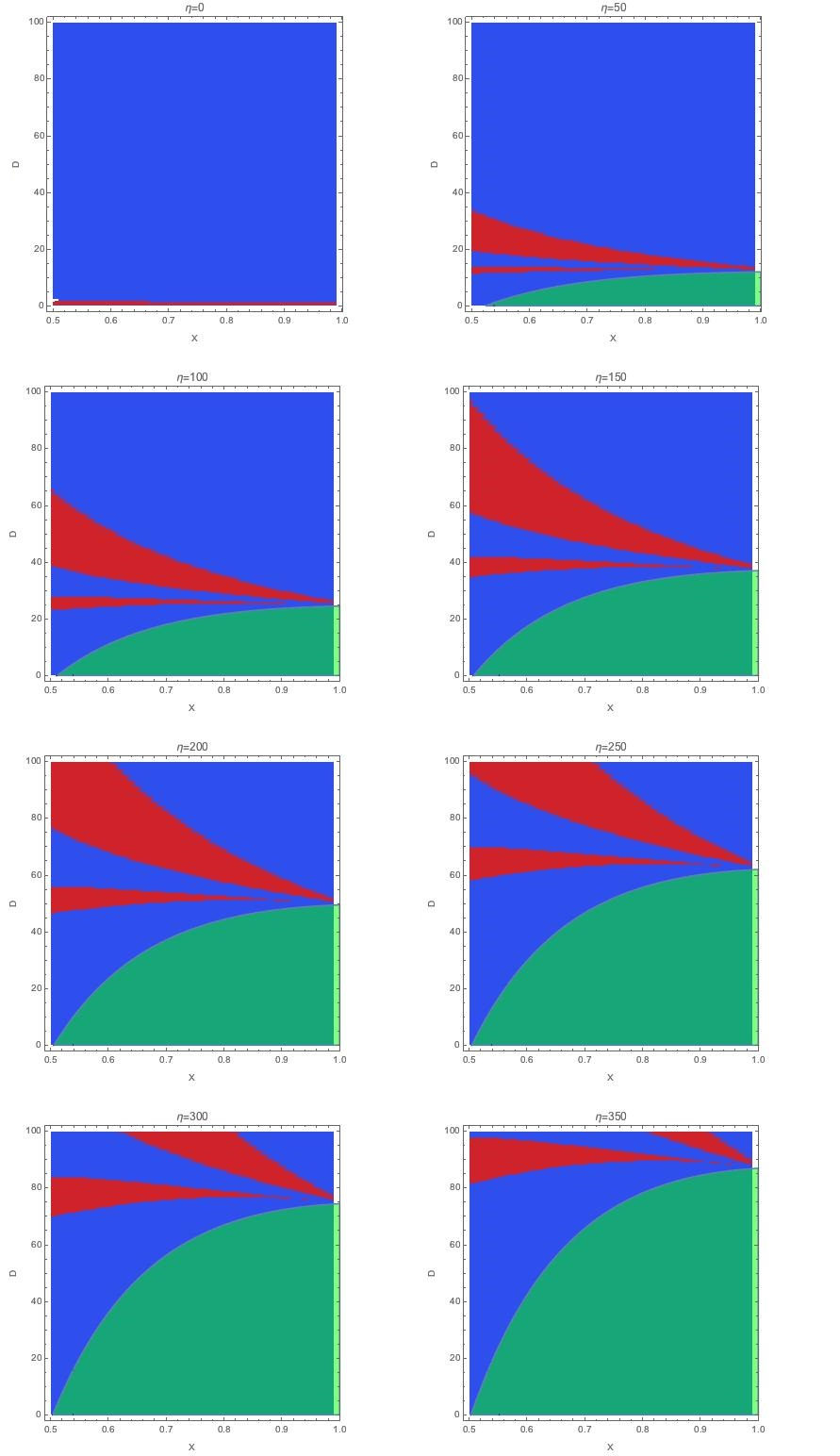}
\caption{\label{eigenfig} Stability analysis for our ansatz. The red region corresponds to parameters that make the real parts of all eigenvalues of $M^{i}_{j}$ negative in the matter era. The green region corresponds to parameters that yield physical values of $c$, $v$, and $\gamma$. As is, the scaling regimes we are predicting are clearly not attractors since the two regions do not overlap for $X<1$.}
\end{figure*}

Instead, our results indicate that the network needs a higher wiggliness in order to survive the more violent energy loss in equilibrium. In fact, this behavior hints at something we will notice when we study the stability of these models: that the wiggly component of our equations leads to instabilities in the scaling regime of these simple models. In other words, the reason our intuition fails us in this analysis is because when we deviate the network from a non-trivial scaling regime it does not generically tend to go back to equilibrium on its own; these scaling regimes are not usually attractors. Also of particular interest is that for these small values of $D$ there appears to be a maximum allowed value of $\mu$ in scaling, $\mu\lesssim 2.2$. This feature disappears if we allow much larger values of this parameter, which however does not seem desirable when we study the stability of the model. Notice also how a slight increase in $D$ seems to dramatically decrease the amount of small-scale structure in any given model (with fixed $\eta$ and $c$) - or, conversely, how it seems to increase the value of $c$ necessary to maintain fixed values of $\mu$ and $\eta$.

The analogous results for $\gamma_{s}$ and $v^{2}_{s}$ can be found in Figs. \ref{fig_matterg} and \ref{fig_matterv}. Naturally, scaling is only allowed for a certain model if it is allowed in all three figures.

We can also carry out a similar analysis for the radiation era ($\lambda=1/2$), whose results for the solutions that come from using the greater roots of Eq.~\ref{eq:varphi_X} are analogous to the ones we have just seen. The results from the other solution, however, are of a much less straightforward interpretation (and are probably of reduced physical significance). If we take a look at the analog of Fig. \ref{fig_matter}, which is Fig. \ref{fig_rad}, this difference is stark: not only is the line corresponding to this new solution of a much different shape and size, but it seems to be extremely insensitive to large variations of $\eta$ while being very sensitive to $D$ (which appears to consistently suppress it).

If we focus instead on the radiation epoch results for $\gamma_{s}$, shown in Fig. \ref{fig_radg}, the situation is even slightly worse: because at some point during our calculations we need to divide very small numbers, our graphs are vulnerable to computational uncertainties. Nevertheless, we are still able to discern a difference in the behavior from the previous case, as well as a robust independence of $\eta$. Without this numerical 'noise', the same kind of differences can be seen in the velocity, which can be found in Fig. \ref{fig_radv}.

\begin{figure*}[h]
\includegraphics[width=3in]{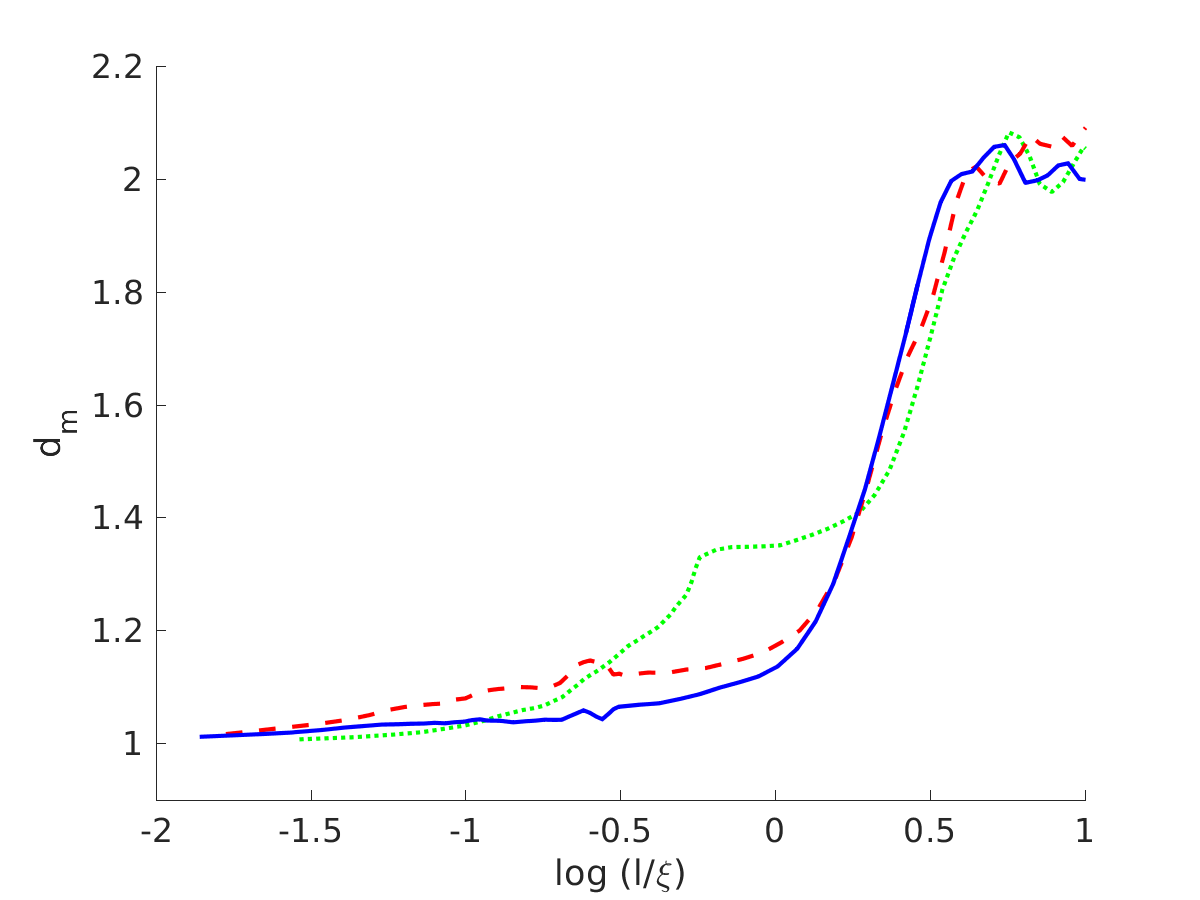}
\includegraphics[width=3in]{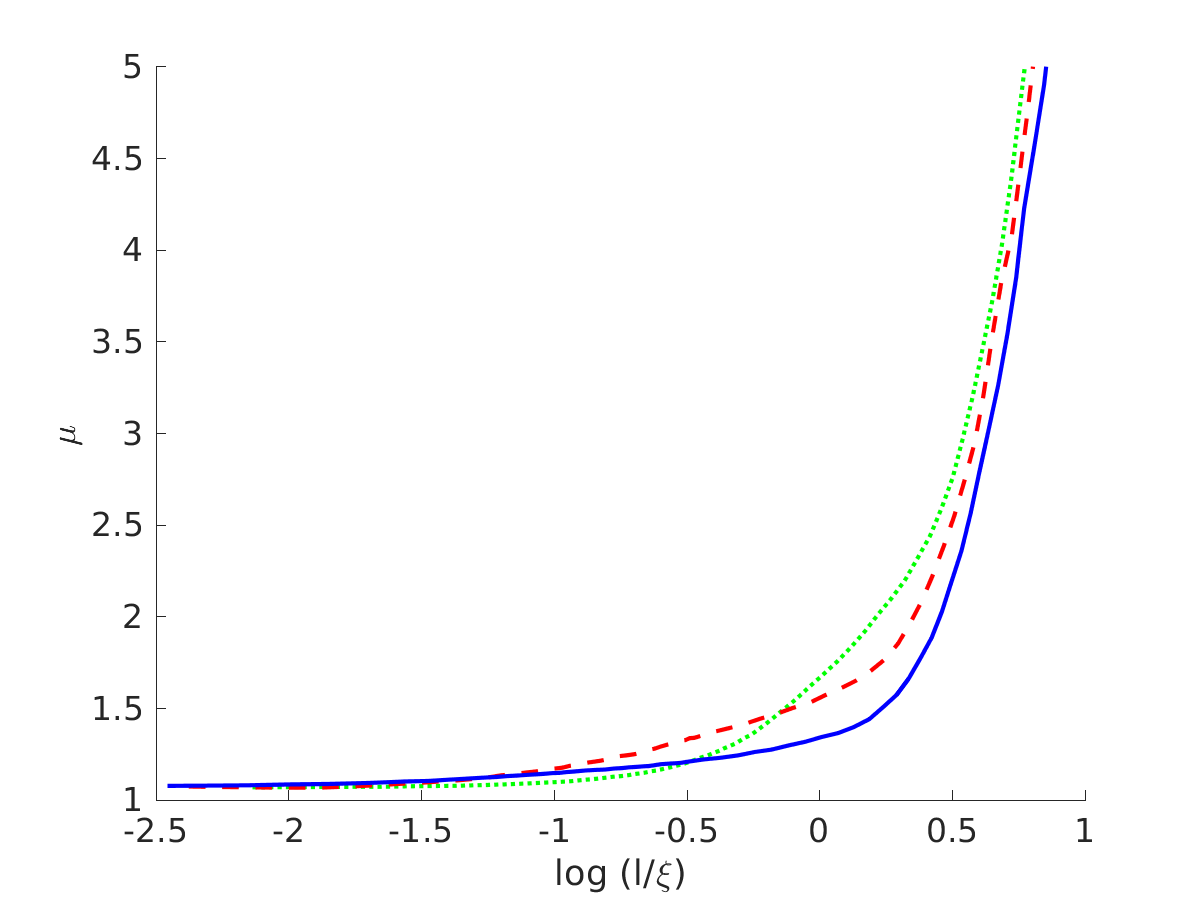}
\includegraphics[width=3in]{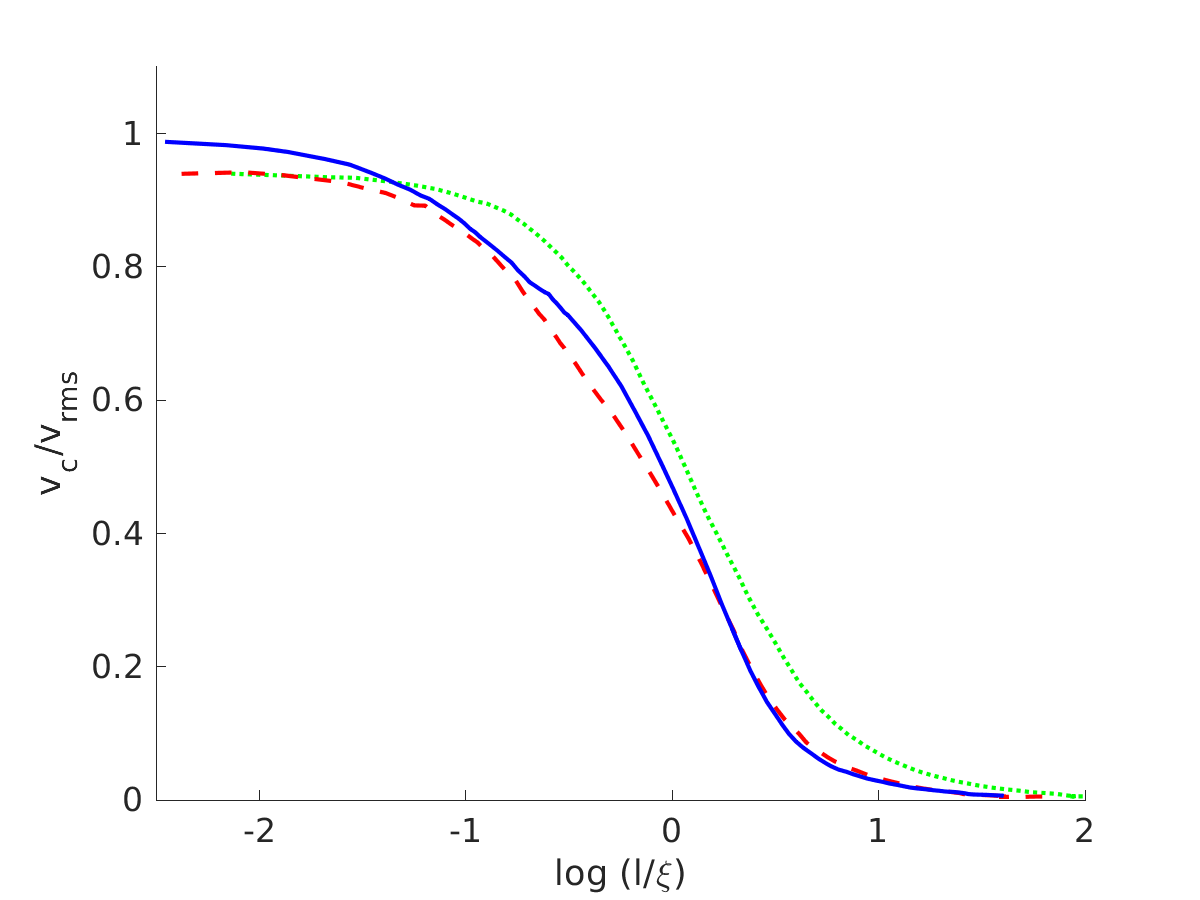}
\caption{\label{wigglyDMV}The behavior of the multifractal dimension, renormalized mass per unit length, and ratio of coherent and RMS velocities as a function of scale, for the final timestep of simulations in flat space (green dotted), radiation era (red dashed) and matter era (blue dotted).}
\end{figure*}

\begin{figure*}[!h]
\includegraphics[width=3in]{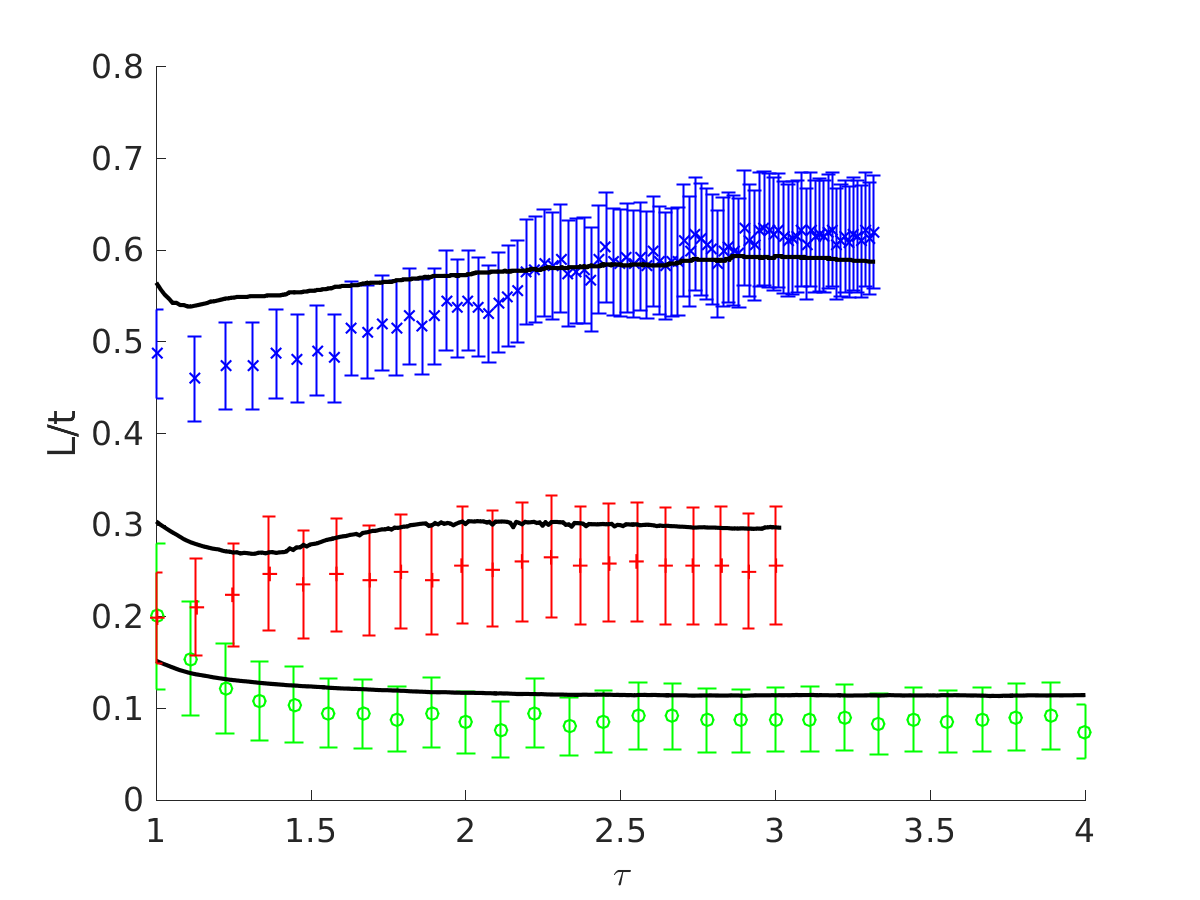}
\caption{\label{wigglyLXI}The behavior of the dimensionless lengthscale $L/t$, calculated from $L=\xi/\sqrt{\mu}$ using the values of $\xi$ and $\mu$ measured directly from the simulation box, in flat space (green data points), radiation era (red) and matter era (blue). Statistical error bars have been estimated from averaging values between neighboring timesteps (hence they are not independent). In all cases the black solid lines depict $L/t$ inferred from the measured total string energy in the simulation box.}
\end{figure*}

\begin{figure*}[!h]
\includegraphics[width=3in]{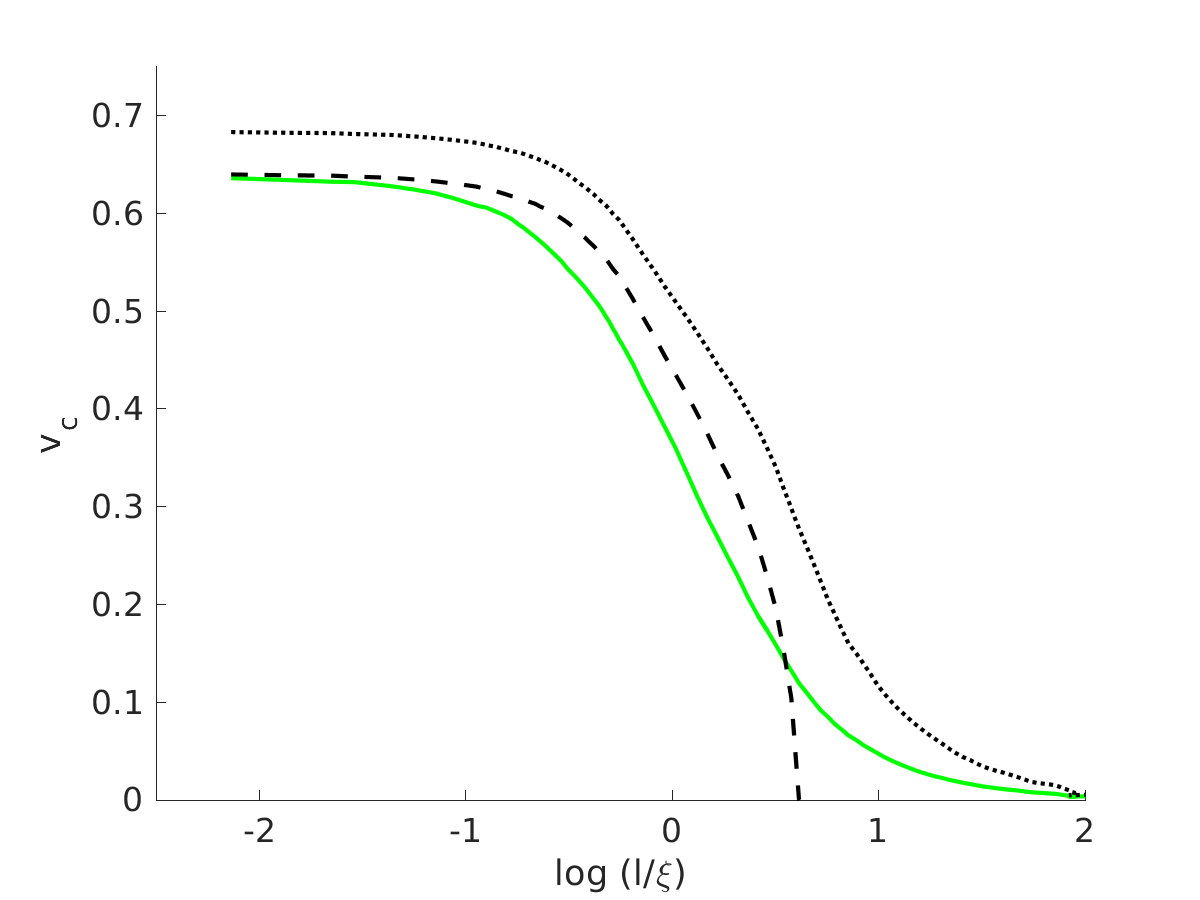}
\includegraphics[width=3in]{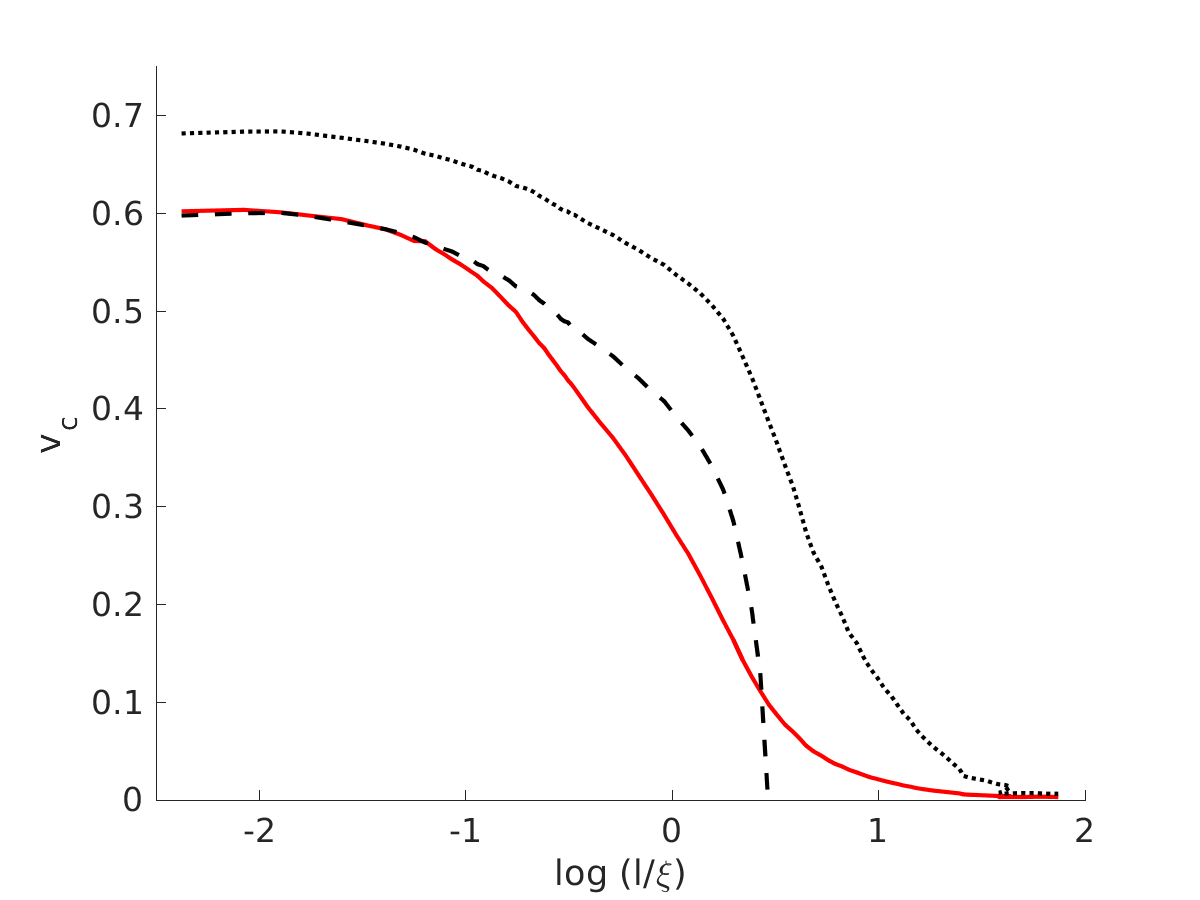}
\includegraphics[width=3in]{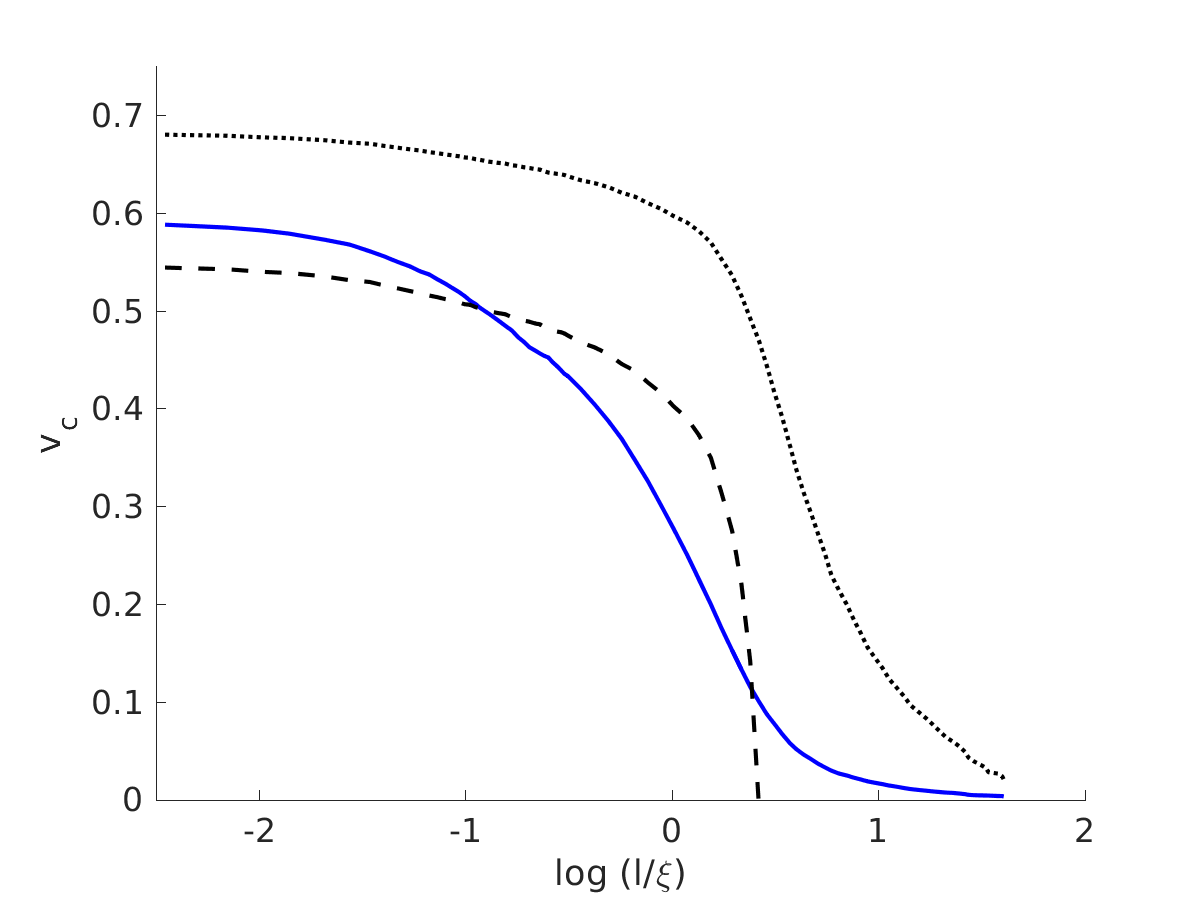}
\caption{\label{wigglyVCOH}The behavior of the average coherent string velocity as a function of scale, for the final timestep of simulations in flat space (solid green), radiation era (solid red) and matter era (solid blue). In all cases the black dashed lines depict the coherent velocity estimated using Eq.~\protect\ref{vrms1} while the black dotted lines depict the one estimated using Eq.~\protect\ref{vvrms}.}
\end{figure*}

\subsection{Exploring (in)stability\label{explore_stable}}

Now that we have found a large family of non-trivial scaling solutions, the time has come to test their stability. We have already mentioned that the shapes we see in Figs. \ref{fig_matter} and \ref{fig_rad} suggest that the introduction of $\mu$ in our equations has spoiled the attractor feature of non-trivial scaling regimes.

Indeed, a direct application of the methodology described in section \ref{stable} reveals that it is not easy (if possible at all) to find stable non-trivial scaling for our heuristic choice of $f$, $f_{0}$, and $s$ as well as our ansatz for $d_{m}\left(\ell\right)$. This difficulty is illustrated in Fig. \ref{eigenfig}.

It should be noted, however, that checking stability requires knowing our energy-loss and multifractal dimension functions with more accuracy than if we just wanted to look for scaling solutions. The reason is that, since $M^{i}_{j}$ depends on derivatives of these functions, second-order corrections can have a first-order impact. As such, what this problem is telling us is not that our ansatze are bad first-order approximations, but rather that we need to go to higher orders if we want to draw conclusions from this sort of stability analysis.

\section{\label{fullev}Comparison with simulations}

\begin{table}
\centering
\begin{tabular}{|c|c|c|c|}
\hline
Parameter & Flat space & Radiation era & Matter era \\
\hline
$L/t$ & 0.10 & 0.27 & 0.62 \\
$v_{\rm rms}$ & 0.65 & 0.64 & 0.59 \\
$\xi/t$ & 0.13 & 0.31 & 0.70 \\
\hline
$\mu(\xi)$ & 1.61 & 1.42 & 1.26 \\
$v_{\ell}(\xi)$ & 0.35 & 0.35 & 0.35 \\
\hline
$\mu(\ell)$ from Eq.~\ref{mu} & 1.69 & 1.32 & 1.27 \\
$v(\ell)$ from Eq.~\ref{vrms1} & 0.38 & 0.50 & 0.44 \\
$v(\ell)$ from Eq.~\ref{vvrms} & 0.51 & 0.60 & 0.62 \\
\hline
\end{tabular}
\caption{\label{table1}Asymptotic values of key network parameters in the simulations of \protect\cite{FRAC}. The first five lines are measured directly from simulations. Although no explicit error bars are provided, they are nominally expected to be around the ten percent level. The last three lines are inferred from the wiggly model, as discussed in the paper.}
\end{table}

Some data from the Goto-Nambu simulations first presented in \cite{FRAC} is shown in Figs. \ref{wigglyDMV}, \ref{wigglyLXI} and \ref{wigglyVCOH}. These are ultra-high resolution simulations, performed in the matter and radiation epochs as well as in flat (Minkowski) spacetime. The initial networks have resolutions of 75 points per correlation length (PPCL), and the simulations subsequently enforce a constant resolution in physical coordinates. Although computationally costly, this is mandatory to obtain accurate diagnostics of the small-scale properties of the network.

Figure \ref{wigglyDMV} shows the scale dependence of key properties for the final timestep of each simulation---respectively we have the multifractal dimension, the renormalized mass per unit length, and the coherent velocity. Note the similarity between the profiles for the different expansion rates (once lengths are re-scaled by the corresponding correlation length $\xi$). As emphasized in \cite{FRAC}, the main difference is the persistence of a significant amount of small-scale structure on scales slightly below the correlation length for the case of Minkowski space. In the expanding universe these structures gradually flow to smaller scales, but this does not happen in the absence of expansion: this interpretation is supported by the fact that on large scales (above the correlation length) the renormalized mass per unit length $\mu$ is larger in Minkowski space than in the expanding case, but the opposite happens for scales below about 1/3 of the correlation length.

Figure \ref{wigglyLXI} compares the values obtained from the simulations for the dimensionless lengthscale $L/t$ in two different ways: calculated from $L=\xi/\sqrt{\mu}$ using the values of $\xi$ and $\mu$ measured directly from the simulation (colored points with error bars, for each of the three epochs), and inferred from the measured total string energy in the simulation box (black line for each case). In the former case, the statistical error bars have been estimated from averaging values between neighboring timesteps (hence they are not independent). We find good overall agreement, although we see that the total string energy diagnostic gives values that are systematically high (though by a small amount) throughout the Minkowski and radiation era simulations as well as early in the matter era one. Is is encouraging that the agreement between the two is much better in the second half of the matter era simulation, where the network is expected to be scaling, as discussed in \cite{FRAC}.

Finally, Fig. \ref{wigglyVCOH} compares the behavior of the average coherent string velocity as a function of scale for the final timestep of each simulations in flat space (solid color lines) to the coherent velocity estimated using Eq.~\ref{vrms1} (solid dashed lines) and using Eq.~\protect\ref{vvrms} (solid dotted lines). One sees that the former provides a good fit on small scales but breaks down (as expected) on scales around three times that of the correlation length (thus, around the scale of the horizon). On the other hand the latter reproduces the overall shape of the curve reasonably well but systematically overestimates its values---by a value which is larger for faster expansion rates.

The asymptotic values of the key network parameters in these simulations are listed in Table \ref{table1}. These can be used for some preliminary calibration of the energy loss terms (which we will do presently), although a full exploration of the parameter space (as was recently done for domain walls \cite{Rybak}) requires additional data that must come from future simulations.

The last three values in Table \ref{table1} are calculated by noting that $\ell$ must be the scale that makes $\xi\left(\ell\right)$ the correlation length. This way $\mu\left(\ell\right)$ is simply given by Eq.~\ref{mu} and can be combined with $v_{rms}$ to yield $v(\ell)$ according to Eq.~\ref{vrms1}. The equivalent result according to Eq.~\ref{vvrms} is included for purely illustration purposes (since, as has been discussed, we do not expect that to apply to these types of simulations).

It is interesting to notice that $\mu\left(\ell\right)$ calculated in this way is compatible with $\mu\left(\xi\right)$ taken directly from the simulations. This could be seen as evidence in favor of the natural identification $\ell=\xi$. Note also that, since $\mu\left(\ell\right)$ must be a non-decreasing function of $\ell$, the central values in Table \ref{table1} actually seem to favor $\ell>\xi$ in flat space and in the matter era. Nevertheless, this counter-intuitive apparent preference should not be too worrying as it is not statistically significant (after all, if $\ell$ truly is just $\xi$, then one would expect this sort of spread where some estimates of $\ell$ are above and some below the correlation length).

There is, however, at least one theoretical consequence of $\ell$ and $\xi$ being at least of the same order, which is that, strictly speaking, we are not working with normal multifractal dimensions, as Eq.~\ref{dm} has
only been shown to hold in the $\ell\ll\xi$ limit \cite{PAP1}. Nevertheless, this has no practical impact on our conclusions
as the simulations we have used to calibrate $d_{m}$ actually probe the left-hand side of Eq.~\ref{dm} rather than the right one.

\begin{figure*}[h]
\includegraphics[scale=0.3]{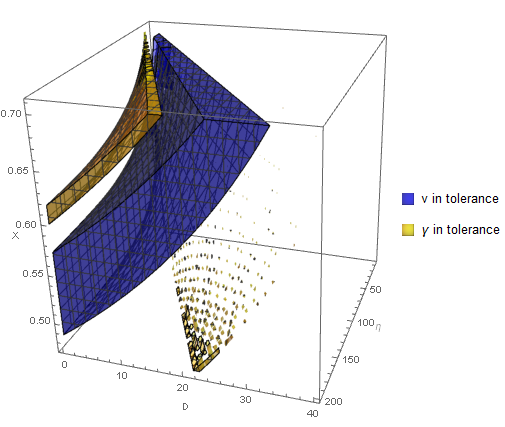} 
\includegraphics[scale=0.3]{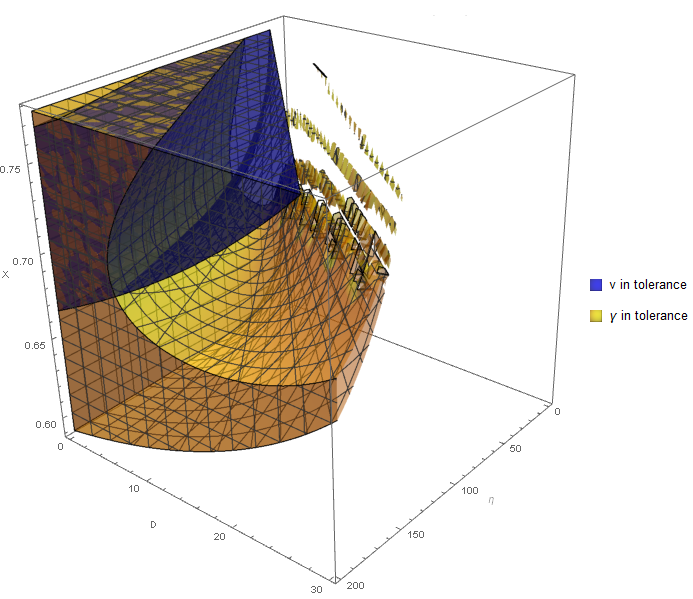}
\caption{\label{3dscaling}Model parameters that allow scaling in agreement with the results in Table \ref{table1} in the radiation era (top) and in the matter era (bottom). The blue region corresponds to scaling velocities allowed by the numerical uncertainty of our simulations and the yellow region is the equivalent for the correlation length. Interestingly, both constraints do not seem to be satisfiable in the radiation era, which seems to confirm our suspicion that strings are not approaching scaling in this era. In the matter era, it appears the overlap between the two regions is not bounded with respect to $\eta$.}
\end{figure*}

The graphs in Fig. \ref{3dscaling} show us which combinations of $\eta$, $D$, and $X_{s}$ (where $X_{s}$ can easily be related to $c$ when the other two are known) admit scaling regimes allowed by the results in Table \ref{table1}. The blue region in this figure corresponds to scaling values of $v\left(\ell\right)$ which are consistent with the values in Table \ref{table1}, and the yellow region is the analogous region concerning the correlation length. As $X$ is plotted in the range allowed by the uncertainty on $\mu\left(\xi\right)$ in the table, the allowed combinations of parameters are those in which the two regions overlap. (There would not be a qualitative difference if we did not use the $\ell=\xi$ identification and instead used the uncertainty on $\mu\left(\ell\right)$.)  These theoretical scaling values were obtained by a simple brute force implementation of the process described in subsection \ref{findscale}. Note also that the scaling regimes depicted here all come from choosing the same root of Eq.~\ref{eq:varphi_X} as the other root yields unphysical values of $c$ in the matter era and too high velocities in the radiation era.

Interestingly, the two colored regions in Fig. \ref{3dscaling} do not overlap in the radiation era, which supports our suspicions that a scaling regime is not being approached in that case. For the matter era, it is curious that allowed combinations of parameters seem to keep existing for arbitrarily large $\eta$ (corresponding to most energy lost to loops being in the form of small-scale loops). 

\section{\label{concl}Conclusions}

With the recent availability of high-quality CMB datasets and the forthcoming availability of comparable gravitational wave datasets, having realistic and accurate models of the evolution of networks of cosmic strings becomes a pressing problem. In this work we have taken further steps towards this goal. Specifically, we have built upon the mathematical formalism described in \cite{PAP1} for a wiggly extension of the VOS model for Goto-Nambu cosmic strings, which can describe the evolution of small-scale structure on string networks, and explored some of the consequences of this model.

Our analysis highlights the fact that the physical nature of the solutions of the model crucially depends on the dominant energy loss mechanisms for the network. Since at present these are still poorly understood, we have introduced a simple ansatz which tackles this problem in the context of an extended velocity-dependent one-scale model. We thus described a general procedure to determine all the scaling solutions admitted by a specific string model and studied their stability, enabling detailed comparisons with future numerical simulations.

Unfortunately, currently available Goto-Nambu and field theory simulations do not yet provide enough information on the small-scale properties of the network to enable a detailed comparison. (Naturally one expects that Got-Nambu simulations will be more useful in this regard, but field theory ones can also play a useful complementary role in the overall calibration of the model's large-scale properties.) The most useful currently available data are those from the Goto-Nambu simulations described in \cite{FRAC}. A comparison of our results with this data supports earlier (more qualitative) evidence that overall scaling of the network is easier to achieve in the matter era than in the radiation era. Still, the fact that a scaling solution can be reached does not {\it per se} ensure that such a solution is stable, and indeed our results show that imposing the requirement that a scaling regime be stable seems to notably constrain the allowed range of energy loss parameters.

In any case, a fully developed model for wiggly cosmic strings is now available. While it has several more free parameters than the original VOS model \cite{MS1,MS2,MS3,MS4}, we emphasize that recent advances in high-performance computing make a detailed calibration of the model's parameters a realistic possibility. Indeed this has been recently done for the analogous model for domain walls \cite{Rybak}, by comparing it to field theory simulations in universes with a range of fixed expansion rates as well as in the radiation-matter transition. In the case of cosmic strings, the possibility of comparing field theory and Goto-Nambu simulations is particularly exciting, both because it will make the calibration process more robust and because it should enable a clearer physical understanding of the relevance of the various energy loss mechanisms.

\begin{acknowledgments}
We are grateful to Ivan Rybak for helpful discussions on the subject of this work.

This work was done in the context of project PTDC/FIS/111725/2009 (FCT, Portugal), with additional support from grant UID/FIS/04434/2013. JV is supported by an STFC studentship. CJM is supported by an FCT Research Professorship, contract reference IF/00064/2012, funded by FCT/MCTES (Portugal) and POPH/FSE (EC).

This work was undertaken on the COSMOS Shared Memory system at DAMTP, University of Cambridge operated on behalf of the STFC DiRAC HPC Facility. This equipment is funded by BIS National E-infrastructure capital grant ST/J005673/1 and STFC grants ST/H008586/1, ST/K00333X/1.
\end{acknowledgments}

\bibliography{paper2}

\end{document}